# Continuum effects for the mean-field and pairing properties of weakly bound nuclei

K. Bennaceur,[1] J. Dobaczewski,[2-4] M. Płoszajczak[1]

[1]*Grand Accélérateur National d'Ions Lourds (GANIL), CEA/DSM – CNRS/IN2P3, BP 5027, F-14021 Caen Cedex, France*
[2]*Institute of Theoretical Physics, Warsaw University, Hoża 69, PL-00681, Warsaw, Poland*
[3]*Department of Physics, University of Tennessee, Knoxville, Tennessee 37996*
[4]*Joint Institute for Heavy Ion Research, Oak Ridge, Tennessee 37831*

## Abstract

Continuum effects in the weakly bound nuclei close to the drip-line are investigated using the analytically soluble Pöschl-Teller-Ginocchio potential. Pairing correlations are studied within the Hartree-Fock-Bogoliubov method. We show that both resonant and non-resonant continuum phase space is active in creating the pairing field. The influence of positive-energy phase space is quantified in terms of localizations of states within the nuclear volume.

Typeset using REVTEX

# I. INTRODUCTION

In the theoretical description of drip-line nuclei, the residual coupling between bound states and the continuum is an essential element of the physical situation. Standard description of many-fermion systems [1] most often invokes the concept of the Fermi sphere of occupied states, and correlations of particles occurring mostly in a narrow zone of the phase space around the Fermi surface. It is obvious that whenever the Fermi energy is close to zero, which by definition is the case in weakly-bound systems, the zone of correlated states must incorporate the phase space of particle continuum.

Among different types of correlations which are important in nuclei, in drip-line nuclei pairing plays a singular role, because the intensity of pairing correlations is a determining factor in establishing position of the last bound nucleus. Indeed, in any isotopic chain, the lightest unbound nucleus corresponds to an odd-$N$ system in which the pairing energy $\Delta$, missing in the binding of the odd neutron, equals to the loss of the mean-field binding when this neutron is shaken off, see e.g. discussion in Refs. [2–4]. Therefore, study of pairing correlations in weakly bound systems is since many years one of the leading subjects of analyses predicting properties of very neutron rich nuclei.

After early analyses, see Refs. [5–11] and references cited and reviewed in Refs. [12,13], the field has now become very active, due to projected developments in radioactive ion beam facilities throughout the world. In particular, methods based on using the Skyrme effective interaction and on the Hartree-Fock-Bogoliubov (HFB) approximation have been applied to both spherical [12–16] and deformed [17–19] drip-line nuclei. Spherical drip-line nuclei have also been studied by using the Gogny interaction and the HFB method [12,16]. In addition, methods based on the relativistic mean field approach [20], with pairing correlations included within the Hartree-Bogoliubov approximation, have been extensively used to study various phenomena in spherical drip-line nuclei [21–28]. Structure of pairs have also been studied by solving the two-body problem exactly, including the continuum effects [29], and the influence of particle resonances on pairing properties have been analyzed in terms of the BCS method [30,31].

In the framework of the mean-field approximation, pairing correlations are consistently included within the variational Hartree-Fock-Bogoliubov (HFB) approach [32]. Such an approach is appropriate for studying bound nuclei, i.e., systems with negative Fermi energy, in which the effects of continuum amount to ensuring that the obtained many-body states are stable with respect to virtual excitations of particles and pairs of particles to positive-energy phase space. In this study we do not consider scattering problems, for which the wave functions are not localized and describe physical situations of particles being scattered off, or emitted by nuclei. On the other hand, bound states, however weakly bound, are always localized and discrete, and such are the basic features of the approach discussed in the present study. A properly executed variational theory, and such is the HFB approach, always yields localized bound states. Moreover, when solved in the coordinate representation, the HFB method takes fully into account all the mean-field effects of coupling to the continuum [6,9,12].

The HFB equations are relatively easy to solve in the matrix representation, i.e., after expanding the quasiparticle wave functions on a suitable basis. When the harmonic oscillator (HO) basis is used, the method is inappropriate for a description of weakly-bound states, because of an incorrect asymptotic properties of the HO wave functions [12]. However, recently developed methods, which use the so-called transformed harmonic oscillator basis [27,33], combine the simplicity of the basis expansion with the corrected asymptotic behavior. In this way, large-scale deformed HFB calculations in drip-line nuclei become recently possible [34]. Another method proposed [35] for a solution of the HFB equations relies on using the natural orbitals; although very promising in principle, it has not been fully implemented in practice yet.

Apart from these attempts, most of the HFB calculations in drip-line systems performed to date use the coordinate representation. Most of these studies have been restricted to spherical symmetry, for which the coordinate-space problem amounts to solving one-dimensional equations. Three-dimensional solutions for deformed nuclei have also been obtained by using hybrid methods of solving the HFB equations in the basis of coordinate-space Hartree-Fock (HF) wave functions [17–19]. All such algorithms use wave functions approximated on spatial grids of points, and the continuum discretized by suitable large-box boundary conditions. Numerical stability and convergence properties of these methods have been thoroughly tested, and are considered to be sufficient for the current mean-field applications, however, more efficient techniques, using the so-called basis-spline Galerkin lattices [36] are also being constructed.

In view of the fact that pairing properties of nuclei near the stability valley are most often treated within the BCS approximation, there have been numerous attempts of using this simpler method in drip-line nuclei too, see e.g. Refs. [37–39,30,31]. Various tests show [6,12,40–42], however, that such an approach is often unstable and/or divergent. A meaningful use of the BCS approximation to the HFB method requires especially careful and conscious treatment, which may, in fact, cost more effort than a straightforward application of the more involved, but certain HFB method itself.

Resonance contribution into the HF+BCS equations was recently studied in Refs. [30,31]. This approximate method can take into account the effect for well separated resonances, when the resonance energies do not depend strongly on the box size or the cutoff procedure. More refined treatments of particle continuum have also recently become available [43,44], although they have not yet been combined with pairing correlations. On the other hand, intuition gained in solving the BCS problem in nuclei tells us that low-energy continuum states must more generally contribute to pairing correlations of drip-line nuclei. In the present paper we aim at quantifying this influence in the framework of the HFB method.

The full-scale HFB calculations and/or the analytically soluble models with realistic potential shapes are invaluable tools for understanding phenomena associated with the continuum in weakly bound systems. In this work, we shall employ the Pöschl-Teller-Ginocchio (PTG) average potential, which is an extension of the textbook Pöschl-Teller potential [45–47], proposed by Ginocchio [48]. It belongs to a broader class of analytically solvable potentials proposed some time ago by Natanzon [49]. The PTG potential has the main features of the nuclear mean field, namely, flat bottom, diffused surface, and asymptotic freedom (i.e., it exponentially vanishes at large distances). For this potential, analytical solutions are available for wave functions and energies of the single-particle bound states and resonances. These advantages make it very useful for applications to nuclear structure problems [50,51].

In Sec. II, we recall the most important features of PTG potential, in particular those concerning the parametrization of the diffuse surface. Sec. III presents some essentials of resonance phenomena and discusses their relation with poles of the $S$-matrix. In Sec. IV, details of the single-particle wave functions and single-particle resonances in the PTG potential are recalled and, in particular, properties of wave function corresponding to weakly bound, virtual and resonance states are discussed. Sec. V contains main results of this work. Here results of the HFB calculations with the PTG input are discussed, and the pairing coupling to single-particle states in the continuum is analyzed. Finally, Secs. VI and VII summarize perspectives of further investigations and the main conclusions we draw from this study, respectively.

## II. PÖSCHL-TELLER-GINOCCHIO POTENTIAL

The PTG potential depends on four parameters: $\Lambda$ which determines its shape and diffuseness, $\nu$ which defines its depth, $s$ (in fm$^{-1}$) which is a scaling factor that can be adjusted to obtain a given mean radius of the potential, and $a$, which allows to account for an effective mass of the particle. In the present study, the effective mass is not discussed, and therefore we take $a=0$, for which the *radial* PTG potential is defined as:

$$V_{PTG}(R) = \frac{\hbar^2 s^2}{2m} \left(v(R) + c(R)\right), \tag{2.1}$$

where $m$ is the mass of free particle, $v(R)$ is the central part,

$$v(R) = -\Lambda^2 \nu_{Lj}(\nu_{Lj}+1)(1-y^2) + \left(\frac{1-\Lambda^2}{4}\right) \tag{2.2}$$
$$\times (1-y^2)\left(2 - (7-\Lambda^2)y^2 + 5(1-\Lambda^2)y^4\right)$$

and $c(R)$ is a (nonstandard) centrifugal barrier,

$$c(R) = L(L+1)\left(\frac{1-y^2}{y^2}\right)(1 + (\Lambda^2-1)y^2). \tag{2.3}$$

Dimensionless variable $\eta$ is proportional to the radial coordinate $R$,

$$\eta = sR, \tag{2.4}$$

and $y$ is implicitly defined by the expression:

$$\eta = \frac{1}{\Lambda^2}\left[\operatorname{arctanh}(y) + \sqrt{\Lambda^2-1}\arctan(y\sqrt{\Lambda^2-1})\right], \tag{2.5}$$

where $0 \leq \eta \leq +\infty$ and, consequently, $0 \leq y \leq 1$.

Parameter $\Lambda$ can take any positive value. (Potentials with $\Lambda<1$, for which in Eq. (2.5) function arctan changes into arctanh, will not be considered in the present study.) In principle, parameters $\Lambda$, $\nu$, and $s$ can take different values for every value of $L$ and $j$. However, below we use $(Lj)$-independent values of shape parameter $\Lambda$ and scaling factor $s$. On the other hand, as indicated in Eq. (2.2), values of depth parameter $\nu$ are independently chosen for each $L$ and $j$, so as to obtain 'reasonable' single-particle spectra, and in particular, to simulate the spin-orbit splitting of single-particle levels which is absent in the PTG potential.

Problem of the nonstandard centrifugal barrier (2.3) deserves a few words of discussion. At the origin ($\eta \to 0$), $y$ decreases as $\eta$, $y \to \eta$, and therefore,

$$s^2 c(R) \longrightarrow \frac{L(L+1)}{R^2} \tag{2.6}$$

becomes the standard centrifugal barrier. On the other hand, at large distances ($\eta \to \infty$), $1-y$ decreases exponentially, i.e.,

$$c(R) \longrightarrow 4\Lambda^2 e^{-2\Lambda^2(\eta-\eta_0)}, \tag{2.7}$$

where

$$\eta_0 = \frac{\sqrt{\Lambda^2-1}}{\Lambda^2} \arctan \sqrt{\Lambda^2-1}, \tag{2.8}$$

and the nonstandard PTG centrifugal barrier exponentially disappears. Consequently, in the PTG potential all partial waves behave asymptotically as the $s$ waves. Therefore, in the following analysis we concentrate on the $L=0$ bound states and resonances for which the centrifugal barrier is absent (see discussion in Secs. IV and V). In some numerical calculations we replace the PTG centrifugal barrier $s^2 c(R)$ (2.3) by the physical barrier $L(L+1)/R^2$, i.e.,

$$V_{PTG'}(R) = \frac{\hbar^2}{2m}\left(s^2 v(R) + \frac{L(L+1)}{R^2}\right). \tag{2.9}$$

Unfortunately, in such a case the analytical solutions do not exist.

For small values of $\Lambda$, the PTG potential is wide and diffuse. In Fig. 1 (lower panel) we present the PTG potentials for $L=0$ and $\Lambda=1$, 3, and 7, with parameters $s$ and $\nu_{Lj}$ (Table VII) chosen in such a way as to keep the depth and radius of the potential fixed. For $\Lambda = 1$ and $\ell=0$ (see Fig. 1), one obtains the Pöschl-Teller potential [45,46] which has been widely studied, e.g., in the molecular physics. For larger values of $\Lambda$, the PTG potential gets steeper and resembles the Woods-Saxon potential. For relatively large values of $\Lambda$ and relatively small values of the depth parameter, one may find a small barrier at the edge of the potential well. Finally, for still larger $\Lambda$, the PTG potential resembles the finite-depth square well potential. The PTG potentials presented in Fig. 1 for $\Lambda=3$ and 7 correspond to the profiles which are interesting from the point of view of simulating the nuclear mean field within a physical range of the diffuseness.

In the middle panel of Fig. 1 we present the $L=4$ *radial* PTG potentials, again for $\Lambda=1$, 3, and 7, and parameters $s$ and $\nu_{Lj}$ given in Table VII. Similarly, the top panel (with the corresponding parameters $s$ and $\nu'_{Lj}$) shows the $L=4$ potentials for the PTG centrifugal barrier $s^2 c(R)$ replaced by the physical centrifugal barrier. One can see that up to a little beyond the minimum, the PTG potential reproduces fairly well the potential with the physical barrier. However, at larger distances, the PTG centrifugal term disappears too quickly.

Nature of the energy eigenstates inside the potential well, as well as the nature of scattering solutions, depend strongly on the shape parameter $\Lambda$. The smaller is the value of $\Lambda$, the broader are the single-particle resonances. Below the critical value of $\Lambda_{\rm crit} = \sqrt{2}$, the resonances entirely disappear, i.e., there are no single-particle resonant states in any partial wave anymore.

## III. RESONANCES

In this study we are interested in properties of the single-particle or quasiparticle resonances in the nuclear average potential. Quantum mechanics deals with numerous types of quasi-stationary states. These states may be classified by the singularities of $S$-matrix or by the mechanism of their production. One-particle shape resonances are perhaps the simplest long-lived states. The particle is captured via quantum tunneling effect to the strongly attractive inner region of the potential. In this way, a quasi-stationary state is formed which, again through the tunneling effect, may leave this inner region. Similar in nature to the one-particle shape resonances are the one-particle virtual states. For these states, the confining barrier is absent but the potential exhibits large jump at the potential boundary region which, in turn, causes jump in the particle wavelength. As a result, the quasi-stationary state is formed, which slowly penetrates from inner to outer region of the potential. The one-particle shape resonances and the one-particle virtual states are specifically quantum phenomena which have no counterpart in the classical physics.

Another kind of resonance states, the so-called Feshbach resonances, are formed when incoming particle excites many particles in the parent nucleus and is captured in the intermediate state for which direct decay channels are closed. The decay of this state is then proceeding through the series of de-excitations of a parent system to either initial channel or to the state of a total system having a lower positive energy [52]. Detailed microscopic theory of these resonances have been worked out in the case, when the coupling of inner excitation to decay channels is weak [53,54]. This has lead to the formulation of continuum shell model (CSM) [54] and to the description, e.g., of giant resonances as quasi-bound $N$-particle states embedded in the continuum [55,56]. The CSM was extended recently to the realistic multi-configurational shell model which describes the coupling between the many body wave functions for bound states and the one-particle scattering continuum. This so-called Shell Model Embedded in the Continuum (SMEC), allows for a simultaneous description not only of the shell model bound states and resonances but also of the radiative capture cross-sections [57,58].

Another type of many-channel quasi-stationary states may result due to the near-threshold singularity [59,60]. This can happen if the overlap of wave function in a given channel with other channel wave functions is small, leading to the effective decrease of the channel-channel coupling and, hence, to the long lifetime. This resonance mechanism is common in the CSM [54,57].

Three-particle, near-threshold long-lived states constitute another class of resonance states which is particularly interesting in the context of drip-line physics. In this class, both near-threshold virtual states ($S$-matrix poles at negative energies on nonphysical sheets) and resonance states ($S$-matrix poles in the complex plane) can be formed. Long-lived states of this class are possible if there exists near-threshold bound, virtual or resonance state in the two-particle subsystem. Multiple transitions of particles between these states of two-particle subsystem are leading to the appearance of effective, long-range exchange forces in the three-particle system [61–63]. An analogous exchange process is also possible in the four-particle systems [64]. The above recollection of most common resonance phenomena in nuclear physics is by no means exhaustive.

## A. Resonances as poles of the S-matrix

In the present Section we recall the standard theory of the S-matrix and introduce the so-called virtual states, which may appear in the single-particle phase space for small energies, and therefore are important for the discussion of pairing correlations in weakly bound systems, see Sec. V.

Poles of the S-matrix can be located in four different regions of the complex $k$-plane, corresponding to four regions of the two-sheet complex energy surface [65] (see Fig. 2). The position of a pole determines both the behavior of the respective wave function and the physical interpretation of the solution. The first region corresponds to the positive imaginary $k$-axis. The wave functions in this region are normalizable, negative energy solutions of the Schrödinger equation and correspond to the bound states of the system. The second region is the negative imaginary $k$-axis. Here solutions of the Schrödinger equation are not normalizable (they are exponentially diverging) and, hence, they are not physical. These solutions correspond to negative energies on the unphysical sheet of the energy surface and they are said to be *virtual* or *antibound* (see Fig. 2c). The third region is the sector between the positive, real $k$-axis and the bisection of the fourth quadrant. Asymptotically, the solutions in this region, which correspond to the resonant states with the complex energy, are oscillating and exponentially decreasing functions. The imaginary part of the energy which is negative in this case, is interpreted as a width of the state. Finally, poles located in the remaining region of the complex $k$-plane, are also said to be virtual.

Figure 2a presents different regions of the complex $k$-plane where the poles of the $S$-matrix are located, and the corresponding regions on the two sheets of the energy surface (Figs. 2b and c). The two sheets are connected along the real positive semiaxis. The arrows in Fig. 2 represent the movement of poles, which results from decreasing the depth of the potential well. In the general case, the poles corresponding to a bound state and to an antibound state move pairwise and cross at a given point (denoted $\kappa$ in Fig. 2a). For $L \neq 0$, this point is situated at the origin. For $L = 0$, $\kappa$ can be found lower on the imaginary axis and the determination of its position is in general not trivial (see [65]). After the crossing, the two poles move pairwise on the lower part of the complex $k$-plane. One considers that they are associated with the resonance phenomenon, once the pole on the right half-plane has crossed the dashed line: $\Re(k) = -\Im(k)$ (see Fig. 2c)), so that one can interpret its complex energy ($\Re(\epsilon) > 0$, $\Im(\epsilon) < 0$) as the energy and the width of a resonance, respectively. Such poles are situated on the unphysical energy sheet but the lower one can influence the positive-energy solution on the real positive energy axis.

Let us now discuss how the poles of the $S$-matrix are related to the widths of resonances. For that, let us consider the short ranged potential $V(R)$ which tends to zero sufficiently fast when $R \to \infty$. The asymptotic ($R \to \infty$) solution of the Schrödinger equation can be written as:

$$\Psi(R \to \infty) = A(k)\exp(-ikR) + B(k)\exp(+ikR)$$
$$\simeq \exp(-ikR) + S(k)\exp(+ikR). \tag{3.1}$$

We will be interested in the poles of $S(k)$ when $A(k)$ vanishes and we shall consider only those poles which are embedded in the fourth quarter ($\Re(k) > 0$, $\Im(k) < 0$) of the complex $k$-plane, i.e., those which are associated with the resonance phenomenon. Near the isolated $i^{th}$ pole, $S(k)$ can be written as:

$$\frac{d(\ln S(k))}{dk} = -\frac{1}{k - k_i}, \tag{3.2}$$

where $k_i$ is the complex pole.

The number of poles in the quarter $\Re(k) > 0$, $\Im(k) < 0$, can be found following the residue theorem:

$$N = \frac{1}{2\pi i}\oint_C \frac{\partial \ln S(k)}{\partial k} dk. \tag{3.3}$$

Consequently,

$$\frac{\partial N}{\partial k} = (2\pi i)^{-1}\frac{\partial \ln S(k)}{\partial k}. \tag{3.4}$$

The density of states:

$$\rho = \left(\frac{\partial N}{\partial k}\right)\left(\frac{\partial k}{\partial \epsilon}\right)^{-1}, \tag{3.5}$$

can be expressed then as follows:

$$\rho = (2\pi i)^{-1}\frac{\partial \ln S(k)}{\partial \epsilon}. \tag{3.6}$$

Inserting (3.2) into (3.6), it is then easy to see that the level density has a local maximum whenever:

$$k = \Re(k_i) \quad , \quad i = 1, \cdots, N \tag{3.7}$$

The value of density at the maximum is:

$$\rho(k = \Re(k_i)) = -\frac{1}{2\pi\Im(\epsilon_i)}, \tag{3.8}$$

and corresponds to the complex energy:

$$\epsilon_i - \epsilon_{th} = \frac{(\hbar k_i)^2}{2m}, \tag{3.9}$$

where $\epsilon_{th}$ is the threshold energy. The density peaks have the Lorentzian shape and the full-width half-maximum of $i^{th}$ peak is given by:

$$\Gamma_i = \frac{1}{\pi\rho(k = \Re(k_i))} = -2\Im(\epsilon_i). \tag{3.10}$$

Following this simple example, we assume that the poles of $S$-matrix on nonphysical energy sheets near the real axis correspond to *almost* all resonance states. Unfortunately, this highly plausible assertion remains only a hypothesis because the relation between resonances and the $S$-matrix poles is not determined so rigorously as the correspondence between the bound states and the $S$-matrix poles on the real axis of the first energy sheet. To affirm the correspondence between the observed resonances and the $S$-matrix poles on nonphysical sheets, certain conditions should be satisfied. First of all, the potential has to be sufficiently analytic and has to fall-off sufficiently rapidly at $R \to \infty$, so that the corresponding $S$-matrix can be safely continued to the unphysical sheets. These conditions are satisfied for the PTG potential, though the examples of potentials where the analytic continuation in the eigenvalue problem brings around the redundant solutions are known as well [66,67]. In general, it is safe to speak about the resonance phenomenon in practical problems if the width is not large, i.e., $\Gamma_{nLj}/\epsilon_{nLj} < 1$ or, in other words, if the distance of the resonance pole from the physical region is small. This latter condition, as we shall show in sect. IV, is never satisfied in the PTG potential for low-lying, near threshold resonances.

## IV. SINGLE-PARTICLE WAVE FUNCTIONS AND RESONANCES IN THE PÖSCHL-TELLER-GINOCCHIO POTENTIAL

### A. Analytical results

Following Ref. [48], we express solutions of the radial Schrödinger equation with the PTG potential (2.1) as functions of the variable $x$:

$$x = \frac{1 - (1 + \Lambda^2)y^2}{1 - (1 - \Lambda^2)y^2}, \tag{4.1}$$

where $y$ is a function of the radial coordinate $R$, given by Eqs. (2.4) and (2.5). In the variable $x$, the Schrödinger equation transforms into the Jacobi equation, and its general solution can be expressed by means of the hypergeometric function:

$$\Psi_{kLj}(R) = \chi_L(R) \left(\frac{1+x}{2}\right)^{\frac{\beta}{2}} \tag{4.2}$$
$$\times F\left(\frac{L+\frac{3}{2}+\beta+\bar{\nu}_{Lj}}{2}, \frac{L+\frac{3}{2}+\beta-\bar{\nu}_{Lj}}{2}, L+\frac{3}{2}; \frac{1-x}{2}\right)$$

where

$$\chi_L(R) = \frac{s^{\frac{1}{2}}}{R}[1 + \Lambda^2 - (1-\Lambda^2)x]^{\frac{1}{4}}\left(\frac{1-x}{2}\right)^{\frac{L+1}{2}}. \tag{4.3}$$

For a given complex momentum $k$, the wave function is specified by two dimensionless parameters:

$$\beta = -\frac{ik}{s\Lambda^2} \tag{4.4}$$

and

$$\bar{\nu}_{Lj} = \left[(\nu_{Lj} + \frac{1}{2})^2 + \beta^2(1 - \Lambda^2)\right]^{\frac{1}{2}}. \tag{4.5}$$

The bound states occur for parameters $\beta$ defined by:

$$\beta = \beta_{nLj}, \tag{4.6}$$

where

$$\Lambda^2 \beta_{nLj} = \left[\left(2n + L + \frac{3}{2}\right)^2 (1 - \Lambda^2) + \Lambda^2 (\nu_{Lj} + \frac{1}{2})^2\right]^{\frac{1}{2}} - \left(2n + L + \frac{3}{2}\right) \tag{4.7}$$

and $n \geq 0$. In this case, the hypergeometric function reduces to the Jacobi polynomial, and the solution reads [48]:

$$\Psi_{nLj} = \mathcal{N}_{nLj}[1 + \Lambda^2 - (1 - \Lambda^2)x]^{\frac{1}{4}}$$
$$\times \left(\frac{1+x}{2}\right)^{\frac{\beta_{nLj}}{2}} \left(\frac{1-x}{2}\right)^{\frac{L+1}{2}} P_n^{(L+\frac{1}{2}, \beta_{nLj})}(x), \tag{4.8}$$

where $\mathcal{N}_{nLj}$ is a normalization factor, and the number of bound states is limited by condition $\beta_{nLj} > 0$, which ensures that the eigenfunctions vanish at the infinity.

The bound-state energies are given by

$$\epsilon_{nLj} = \frac{\hbar^2 s^2}{2m} \mathcal{E}_{nLj}, \tag{4.9}$$

where the dimensionless eigenenergies are

$$\mathcal{E}_{nLj} = -\Lambda^4 \beta_{nLj}^2. \tag{4.10}$$

It can be seen from the above equation that the tail of a bound-state wave functions for $\eta \to \infty$,

$$\Psi_{nLj}(R) \propto e^{-\beta_{nLj}\Lambda^2(\eta - \eta_0)}, \tag{4.11}$$

does not explicitly depend on $L$, i.e., there is no influence of centrifugal barrier on any partial wave.

If $\Lambda^2 > 2$, the resonances will occur for integer $n$ such that:

$$n > \frac{1}{2}[\Lambda(\Lambda^2 - 2)^{-1/2}(\nu_{Lj} + \frac{1}{2}) - 1] \tag{4.12}$$

The dimensionless resonance energy is then:

$$\mathcal{E}_{nLj} = (2n + L + \frac{3}{2})^2(\Lambda^2 - 2) - \Lambda^2(\nu_{Lj} + \frac{1}{2})^2, \tag{4.13}$$

and the dimensionless resonance width is:

$$\gamma_{nLj} = 4\left(2n + L + \frac{3}{2}\right) \times \left[(2n + L + \frac{3}{2})^2\left(\Lambda^2 - 1\right)\right.$$
$$\left. - \Lambda^2\left(\nu_{Lj} + \frac{1}{2}\right)^2\right]^{\frac{1}{2}}. \tag{4.14}$$

In the limit: $\mathcal{E}_{nLj} \to 0$, the resonance width is:

$$\gamma_{nLj} \to \gamma_{nLj}^{(0)} = 4(2n + L + \frac{3}{2})^2, \tag{4.15}$$

and, hence, the ratio: $\gamma_{nLj}/\mathcal{E}_{nLj} \to \infty$, for all values of parameter $\Lambda$. For small $\mathcal{E}_{nLj}$, we have:

$$\frac{\gamma_{nLj}}{\mathcal{E}_{nLj}} = \frac{\gamma_{nLj}^{(0)}}{\mathcal{E}_{nLj}} + 2 - 4\frac{\mathcal{E}_{nLj}}{\gamma_{nLj}^{(0)}}. \tag{4.16}$$

This ratio depends strongly on $\Lambda$. For large $(n, L)$, the quantities $\mathcal{E}_{nLj}$ and $\gamma_{nLj}$ are proportional to $(2n + L)^2$ and their ratio is:

$$\frac{\gamma_{nLj}}{\mathcal{E}_{nLj}} = 4\frac{(\Lambda^2 - 1)^{1/2}}{\Lambda^2 - 2}, \tag{4.17}$$

i.e. $\gamma_{nLj}/\mathcal{E}_{nLj} < 1$ for $\Lambda > (10 + 4\sqrt{5})^{1/2}$. Therefore, in our numerical examples we present results for $\Lambda=3$ and 7, which are the values on two sides of the limiting case defined by the widths of resonances being equal to resonance energies.

Solutions in the continuum can be found by analytically continuing the eigenfunctions from a discrete negative energy to positive continuous energy (see Eq. (4.2)). The solutions obtained in this way are proportional to hypergeometric functions. Imposing the boundary condition that an incoming wave has the momentum $k$, one can determine the scattering function for each angular momentum [48]. Using the asymptotic behavior of the most general solution (4.2) one obtains the matrix elements of the $S$-matrix:

$$S(k) = (-1)^{L+1} \exp\left(2\beta[\Lambda^2\eta_0 + \ln\Lambda]\right)$$
$$\times \frac{\Gamma[-\beta]\Gamma[(L + \frac{3}{2} + \beta + \bar{\nu}_{Lj})/2]}{\Gamma[\beta]\Gamma[(L + \frac{3}{2} - \beta + \bar{\nu}_{Lj})/2]} \tag{4.18}$$
$$\times \frac{\Gamma[(L + \frac{3}{2} + \beta - \bar{\nu}_{Lj})/2]}{\Gamma[(L + \frac{3}{2} - \beta - \bar{\nu}_{Lj})/2]}.$$

Expression (4.18) yields the matrix elements of $S$-matrix for the Schrödinger problem with the PTG potential without any restrictions, i.e., (4.18) contains informations about all mathematical solutions, both physical and unphysical ones. The poles of the $S$-matrix in the variable $k = is\Lambda^2\beta$, correspond to the *remarkable solutions* [65] depending on the asymptotic behavior of solutions.

## B. Wave functions

In the following we aim at analyzing the influence of weakly bound states and low-energy resonances on pairing properties of nuclei near neutron drip lines. Therefore, knowing the analytical solutions available for the PTG potential, we chose three sets of parameters such that the $3s_{1/2}$ state is either weakly bound, or virtual, or there is a low-lying resonance in the $s_{1/2}$ channel. These three physical situations can be achieved by fixing parameters $\Lambda$ and $s$ and shifting the depth parameter $\nu_{s\frac{1}{2}}$. In the specific examples discussed below, we use the parameters listed in Table II, where also the corresponding energies are given.

Wave functions of the resonant, virtual, and bound $3s_{1/2}$ states are shown in Fig. 3. The resonant wave function is calculated at the energy equal to the real part of the pole shown in Table II. Normalizations of the virtual and resonant wave functions are chosen so as to match the height of the first maximum of the bound wave function. The three wave functions illustrate properties of single-particle phase space in the situation where the $3s_{1/2}$ state leaves the realm of bound states, dwells for a short time in a ghost-like zone of virtual states, and then reappears as a decent (although very broad) resonance. From the point of view of the asymptotic properties, pertinent to the scattering problems, these three wave functions are completely different. On the other hand, from the point of view of their structure inside the nucleus (see the inset in Fig. 3), they are almost exactly identical. In fact, in the scale of the inset, the resonant wave function cannot be distinguished from the bound one.

In Sec. V we also study several cases of different positions of the $2d_{3/2}$ states. In particular, we use two values of the depth parameter $\nu_{d\frac{3}{2}}$, which are listed in Table II together with the corresponding energies of the $2d_{3/2}$ PTG resonance and virtual states. Whenever the spectrum in all partial waves is required, like in the HFB calculations below, we combine $\nu_{d\frac{3}{2}}$=4.900 with all the three values of $\nu_{s\frac{1}{2}}$, and $\nu_{s\frac{1}{2}}$=5.034 with all values of $\nu_{d\frac{3}{2}}$.

## C. Localizations and phase shifts

Since the pairing potentials depend self-consistently on the pairing densities, they are concentrated inside the nucleus, i.e., they are nonzero inside, and go to zero outside the nuclear volume. Therefore, the more the continuum wave functions are concentrated inside the nucleus, the more they are sensitive to the pairing coupling. In order to quantitatively characterize the wave functions from this point of view, we introduce their 'localization' as the norm inside the sphere of radius $R_{\text{Loc}}$,

$$L[\psi] = \int_0^{R_{\text{Loc}}} |\psi(r)|^2 \, dr, \tag{4.19}$$

where the radius $R_{\text{Loc}}$ is, for the purpose of the present study, arbitrarily fixed at $R_{\text{Loc}} = 1.5 \times R_{1/2} = 7.578 \, \text{fm}$, while $R_{1/2}$ is the radius where the potential drops to its half value. [Note that the volume element $4\pi r^2$ is included in the definition of wave function $\psi(r)$.] For the bound states, localization is just the probability to find the particle inside the sphere $R < R_{\text{Loc}}$. For the continuum wave functions which are not normalizable, localization depends on the chosen normalization condition.

Below we present results for the continuum wave functions normalized in two ways. Firstly, we may normalize them to unity inside the box of a large radius. (The value of $R_{\text{box}}$=30 fm is chosen for all results obtained in the present study.) Outside the box, the wave functions oscillate and have an infinite norm. In this case, the localization is a fraction of the probability to find the particle inside the nucleus relative to the probability to find it inside the box. Absolute values of such a localization depend, of course, on the radius of the box, however, we are only interested in comparing relative localizations of the wave functions with different energies. Secondly, we may normalize the continuum wave functions in such a way that they all have a common arbitrary amplitude in the asymptotic region. For wave functions in which the volume element $4\pi r^2$ is included, as is the case here, their amplitudes do not asymptotically depend on $r$. Again, the absolute values of this localization depend on the value of the common amplitude, but the relative values tell us at which energy the given wave function is better concentrated inside the nuclear volume. One may chose other normalization conditions (to a delta function in energy or in momentum, for example), but we did not find any advantage in looking at the corresponding localizations, and we discuss here only the two ways described above.

Fig. 4a shows values of localizations of the $s_{1/2}$ PTG continuum states, calculated analytically, for the normalization in the box (solid line) and for the normalization to a constant amplitude (dashed line). Calculations are performed for the PTG potential for which there is a weakly bound $3s_{1/2}$ state (cf. Table II), and therefore, the lowest resonance in the $s_{1/2}$ channel appears high in energy. The wide bump in the localization, which is seen near 40 MeV reflects the known fact that the $S$-matrix pole at $(39-18i)$ MeV generates continuum states which are more localized than the continuum states far from the resonant energies. However, values of localizations near the maximum are only a factor of 2-3 larger than those far from the maximum, i.e., one cannot *à priori* expect that the only continuum states which can couple to the pairing field are those close to resonances.

One can see that the localization obtained from normalizing the continuum states in the box (solid line) presents numerous wiggles, which appear when the consecutive half-waves enter the box. On the other hand, the localization obtained from the amplitude normalization is given by a smooth curve. Both localizations are fairly similar, and therefore, different normalization prescriptions do not affect our conclusion about the relative localizations of the continuum states.

There is some difference between these two different normalization prescriptions when $\epsilon \to 0$. At the origin, $\phi(r) = A \sin(kR)$. In case of the normalization in the box, we can assume for $k \to 0$ that this expression is valid in the whole box. Then : $\int_0^{R_{\text{Loc}}} \phi^2(r) dr / \int_0^R \phi^2(r) dr \simeq \int_0^{R_{\text{Loc}}} r^2 dr / \int_0^R r^2 dr = (R_{\text{Loc}}/R)^3 = $ const. We have verified this assertion also numerically. In the case of normalizing to an amplitude $A$, for $k \to 0$ one gets: $\int_0^{R_{\text{Loc}}} \phi^2(r) dr \simeq |A|^2 k^2 r^3 / 3 \longrightarrow 0$. Therefore, the dashed curve turns sharply down when approaching $\epsilon$=0.

In Fig. 4a we also present localizations of the continuum states calculated numerically (circles) in the same box of $R_{\text{box}}$=30 fm, and using the discretization of wave functions on a mesh of nodes equally spaced by $H$=0.25 fm. One can see that the numerical results perfectly reproduce the analytical calculations for the same normalization. In the numerical calculations, the box plays merely a role of selecting from the infinite continuous set of positive-energy solutions a discrete subset of wave functions which vanish at the box boundary. Apart from that, the numerically calculated wave functions are very precise representations of the exact wave functions for some specific discretized values of the energy.

Fig. 4b shows the phase shifts of the same continuum states calculated analytically (solid line) and numerically (circles). (In this case, normalization of the continuum wave functions does not play any role.) Again, one can see that the numerical results very precisely reproduce the analytical ones in the whole range of studied energies.

Usually, one identifies the resonance when the phase shift passes $\pi/2$ (or better $n\pi/2$, as the phase shift is defined modulo $2\pi$). This definition works well for narrow, well separated resonances. In the case of PTG potential, the resonances are broad and this definition is inadequate. It is then better to identify the resonance with the inflection point in the derivative $d\delta(\epsilon)/d\epsilon$. This also demonstrates real difficulty in identifying resonances in potentials with diffuse surfaces and proves again the advantage of the soluble models where the $S$-matrix is analytically known and the poles can be analytically studied.

## V. PAIRING IN WEAKLY BOUND SYSTEMS

A variational mean-field approach to pairing correlations results in the HFB equations [32]. In weakly bound systems, these equations should be solved in coordinate space in order to properly take into account the closeness of the particle continuum [6,12]. In the most general non-local coordinate representation, the HFB equations have the form of the following matrix integral eigenequation:

$$\int d^3 \boldsymbol{r}' \sum_{\sigma'} \begin{pmatrix} h(\boldsymbol{r}\sigma, \boldsymbol{r}'\sigma') & \tilde{h}(\boldsymbol{r}\sigma, \boldsymbol{r}'\sigma') \\ \tilde{h}(\boldsymbol{r}\sigma, \boldsymbol{r}'\sigma') & -h(\boldsymbol{r}\sigma, \boldsymbol{r}'\sigma') \end{pmatrix} \begin{pmatrix} \phi_1(E, \boldsymbol{r}'\sigma') \\ \phi_2(E, \boldsymbol{r}'\sigma') \end{pmatrix}$$
$$= \begin{pmatrix} E + \lambda & 0 \\ 0 & E - \lambda \end{pmatrix} \begin{pmatrix} \phi_1(E, \boldsymbol{r}\sigma) \\ \phi_2(E, \boldsymbol{r}\sigma) \end{pmatrix}, \tag{5.1}$$

where $E$ is the quasiparticle energy, $\lambda$ is the Fermi energy, and $h(\boldsymbol{r}\sigma, \boldsymbol{r}'\sigma')$ and $\tilde{h}(\boldsymbol{r}\sigma, \boldsymbol{r}'\sigma')$ are the mean-field particle-hole (p-h) and particle-particle (p-p) Hamiltonians, respectively. Contrary to the HF equations, which define one-component (single-particle) wave functions [the eigenstates of $h(\boldsymbol{r}\sigma, \boldsymbol{r}'\sigma')$], the HFB method gives two-component (quasiparticle) wave functions (the upper and lower components are denoted by $\phi_1(E, \boldsymbol{r}\sigma)$ and $\phi_2(E, \boldsymbol{r}\sigma)$, respectively).

In the following, we solve the HFB equations (5.1) by fixing the p-h Hamiltonian to be equal to the sum of the kinetic energy (with constant nucleon mass) and PTG' potential (2.9). In this way, we study self-consistency only in the pairing channel, while the single-particle properties are kept unchanged, and under control. For example, the single-particle energies and resonances do not change during the HFB iteration, and are not affected by the pairing properties, which would have not been the case had we allowed the usual HFB coupling of the p-h and p-p channels. Moreover, within such an approach we only need to solve the HFB equations for neutrons, i.e., for the particles which exhibit the weak binding under study here. Note that for the physical centrifugal barrier included in all the $L>0$ partial waves, the energies of single-particle bound states and resonances are not given by analytical expressions (4.10) and (4.13). However, the barrier does not appear in the $s_{1/2}$ channel, and these energies are still given analytically.

Two parameters of the PTG potential have been fixed at values used in the previous Sections, namely, $\Lambda=7$ and $s=0.04059$, while the depths parameters $\nu_{Lj}$ (Tables II and III) have been chosen in such a way [48] as to simulate a hypothetical single-particle neutron spectrum in drip-line nuclei with $N \lesssim 82$. For the scope of the present study, details of this spectrum are insignificant; we only aim at realizing the physical situation where the PTG' $3s_{1/2}$ or $2d_{3/2}$ states are near the threshold (close to zero binding energy) and at the same time the Fermi energy is negative and small.

Contrary to the p-h channel, the full self-consistency is required in the p-p channel, with the p-p Hamiltonian given by the local pairing potential [6,12]:

$$\tilde{h}(\bm{r}\sigma, \bm{r}'\sigma') = \tilde{U}(\bm{r})\delta(\bm{r}-\bm{r}')\delta_{\sigma,\sigma'}, \tag{5.2}$$

where

$$\tilde{U}(\bm{r}) = \frac{1}{2}V_0\tilde{\rho}(\bm{r}), \tag{5.3}$$

and

$$\tilde{\rho} = -\sum_{0<E_n<E_{\max}}\sum_{\sigma}\phi_2(E_n,\bm{r}\sigma)\phi_1^*(E_n,\bm{r}\sigma). \tag{5.4}$$

Potential (5.3) corresponds to the pairing force given by the zero-range interaction,

$$V(\bm{r}_1-\bm{r}_2) = V_0\delta(\bm{r}_1-\bm{r}_2). \tag{5.5}$$

Strength parameter $V_0$ has been arbitrarily fixed at $V_0=-175\,\text{MeV}\,\text{fm}^3$. The pairing phase space, given by the cut-off parameter $E_{\max}$, has been fixed according to the prescription formulated in Ref. [6]. In Eq. (5.4) we used the fact that in order to discretize the continuum HFB states, the HFB equation is solved in a suitable spatial box. The HFB results presented below have been obtained with the same box size of $R_{\text{box}}=30\,\text{fm}$ as those discussed in Sec. IV C.

As seen from Eqs. (5.1)–(5.3), the intensity of the pairing coupling [i.e., the off-diagonal term in Eq. (5.1)] is given by the integral of the wave functions with the pairing density $\tilde{\rho}(\bm{r})$. This integral can be approximated in the following way:

$$\int d^3\bm{r}\sum_{\sigma}\phi_1^*(E_n,\bm{r}\sigma)\tilde{\rho}(\bm{r})\phi_2(E_n,\bm{r}\sigma) \simeq \tilde{\rho}_0\left(N_n L[\phi_1^*]\right)^{1/2}, \tag{5.6}$$

where $L[\phi_1^*]$ is the localization of the upper HFB wave function, defined as in Eq. (4.19), and $N_n$ is the norm of the lower HFB wave function:

$$N_n = \int d^3\bm{r}\sum_{\sigma}|\phi_2(E_n,\bm{r}\sigma)|^2. \tag{5.7}$$

Equation (5.6) gives only a very crude approximation, which aims only at showing the main trends. It is based on two assumptions: (i) that the pairing potential is constant within the radius $R_{\text{Loc}}$ of the sphere for which the localization is defined, and zero otherwise. Zero-range pairing force (5.5) leads to the volume-type pairing correlations [12], for which the pairing densities are spread throughout the nucleus, and can be crudely approximated by a constant value $\tilde{\rho}_0$. Another assumption is: (ii) that the lower and upper HFB wave functions are proportional to one another within the radius of $R_{\text{Loc}}$. We know that this assumption holds only in the BCS approximation, while in the HFB approach the lower and upper components are different, including different nodal structure [12]. For all energies $E_n$, the lower components are localized inside the nucleus, and their norms $N_n$ give contributions to the particle number (see examples of numerical values presented in Ref. [12] and in the following subsections). On the other hand, the continuum upper HFB wave functions behave asymptotically as plane waves, however, their pairing coupling is dictated by their localizations. We are not going to use Eq. (5.6) in any quantitative way; we only use it as a motivation to look at localizations of the upper HFB components as measures of how strongly given continuum states contribute to pairing correlations.

### A. The $s_{1/2}$ continuum

Figures 5 and 6 show the $s_{1/2}$ localizations and phase shifts, respectively, of the upper HFB quasiparticle wave functions calculated numerically (dots), compared with the analytically calculated localizations and phase shifts of the PTG states (lines). We are interested in the low-energy $s_{1/2}$ continuum, and therefore the Figures show results well below the broad $(39-18i)$ MeV resonance discussed in Sec. IV C. The PTG results (no pairing) are shown as functions of the single-particle energy $\epsilon$, while the HFB results are plotted as functions of the quasiparticle energy $E$ shifted by the Fermi energy $\lambda$, i.e., $E+\lambda$. In this way, at large energies the paired and unpaired energy scales coincide.

Three panels presented in Figs. 5 and 6 correspond to the PTG $3s_{1/2}$ states being resonant (a), virtual (b), and weakly bound (c), cf. Table II. In order to realize these three different physical situations, the bottom of the PTG potential has to be shifted by about 2.75 MeV. This illustrates the "width" of the zone where the $3s_{1/2}$ states are virtual. Apparent shift in the single-particle energies is much smaller; the $3s_{1/2}$ state moves then from the $-74$ keV bound-state energy to the 74 keV resonance energy (Table II). Even if the shift in the single-particle states is so small, the corresponding canonical states move by about 1.75 MeV (Table IV). These latter states appear to be entirely unaffected by the dramatically different character of the single-particle $3s_{1/2}$ states (Fig. 3). Their wave functions, shown in Fig. 7, are almost identical in the three cases (a)-(c).

Since the canonical states govern the pairing properties of the system, see discussion in Ref. [32], their positions explain the changes in the overall pairing intensity $\langle\Delta_N\rangle$, and in the $3s_{1/2}$ occupation $v_{\text{can}}^2$, shown in Table IV. It is worth noting that shifting the canonical $3s_{1/2}$ level from its $\sim$0.8 MeV distance to the Fermi level to a distance of $\sim$2.5 MeV decreases the average pairing gap by as much as about 300 keV.

Apart from the decrease of the pairing intensity described above, there is no other qualitative change in pairing properties when the $3s_{1/2}$ state becomes unbound. The role of the weakly bound state is simply taken over by the low-energy continuum. The presence and position of the low-energy $s_{1/2}$ resonance is not essential for the pairing properties. In particular, it would have been entirely inappropriate to use this resonance, and its energy of 74 keV, in any approximate scheme in which the full continuum was replaced by the resonances only.

In all the three cases (a)-(c), in the region of energies between 10 and 20 MeV (Fig. 5) one can see the "background" localization of the order of 0.2. These energies are far away from resonances, and therefore illustrate localizations of a "typical" non-resonant continuum. Near the resonances (see Figs. 4a and 5a), localizations are larger, up to 0.3, but do not at all dominate over the background value. Therefore, based on estimate (5.6) one may expect that the pairing coupling of the resonant and non-resonant $s_{1/2}$ continuum is fairly similar.

The magnitude of this coupling seems to depend primarily on the $s_{1/2}$ single-particle localization in the 2–3 MeV zone above the Fermi surface. (Note that cases (a)-(c) differ only by positions of the $s_{1/2}$ states; the spectra in other partial waves are kept unchanged.) The weakly bound $3s_{1/2}$ PTG state generates large continuum localization right above the threshold, and the opposite is true for the low-lying $3s_{1/2}$ PTG resonance. Therefore, the average pairing gap $\langle \Delta_N \rangle$ is substantially larger in case (c) than in case (a), and as a consequence, the HFB localizations in case (c) differ strongly from the PTG localizations, while in case (a) they are almost identical. The same pattern clearly also appears for the HFB phase shifts, as compared to the PTG phase shifts, Fig. 6.

Figure 8 shows norms $N_n$ of the lower HFB components, Eq. (5.7). Since the lower HFB components are not mutually orthogonal, $N_n$ cannot be associated with the occupation probabilities. On the other hand, the canonical occupation factors $v_{\text{can}}^2$ do play such a role, because the canonical states form an orthogonal basis. Comparing values of $N_n$ (Fig. 8) and $v_{\text{can}}^2$ (Table IV), one sees that the canonical $3s_{1/2}$ state collects all the occupation strength of the quasiparticle states in the low-energy continuum. Of course, numbers $N_n$ scale with the overall pairing strength, i.e., they are smaller in case (c) than in case (a), but in every case all quasiparticle states below about 5 MeV significantly contribute to $v_{\text{can}}^2$. Moreover, choosing only one quasiparticle state in this zone (call it resonance or not), would have provided only for about one third of the occupation factor. This illustrates again that the $s_{1/2}$ continuum has to be taken as a real continuum (discretized, if needed), and not through any single representative.

## B. The $d_{3/2}$ continuum

Since the low-energy $s_{1/2}$ resonances are always very broad, the distinction between the resonant and non-resonant continuum is not very clear in this channel. On the other hand, pairing coupling of very narrow high-$j$ resonances is not very different from that of any other bound states, and hence they are not very interesting to look at. Therefore, in addition to investigating the $s_{1/2}$ resonances, we also study here an intermediate case of low-energy $2d_{3/2}$ resonances, which can be narrow or broad depending on their positions inside the centrifugal barrier. For that we use the PTG' potential which contains the physical centrifugal barrier, Eq. (2.9), and chose several different values of the depth parameter $\nu_{d\frac{3}{2}}$.

First, we study two cases listed in Table II, chosen in such a way that the resonances are located at two different positions deep inside the centrifugal barrier. (For the chosen depths, in the PTG potential the $2d_{3/2}$ states are resonant (d) or virtual (e); their analytical energies are given in Table II.) Although for the physical barrier the analytical results are not available, one can estimate the resonance energies to be about $\epsilon_{\text{res}} \simeq (1.5-0.4i)$ MeV (d) and $\epsilon_{\text{res}} \simeq (0.9-0.2i)$ MeV (c), respectively. In the two cases, the resonances are located at about 3.5 MeV and 4.1 MeV below the top of the barrier which is about 5 MeV high.

One can see that the inclusion of the physical barrier moves the PTG resonance (Table II) up in energy and decreases its width by a factor of eight, while the PTG virtual state is transformed into a true, rather narrow resonance.

Figure 9 shows the HFB and PTG' localizations of the $d_{3/2}$ continuum states. Of course, the low-energy PTG' resonances create narrow peaks of localizations at the resonant energies, and having resonance widths. Insets in Fig. 9 show that the pairing correlations shift the PTG' localizations to slightly higher energies, but otherwise the HFB and PTG' localizations are very similar. Beyond the narrow resonances, the HFB and PTG' localizations reach the non-resonant background values of about 0.3–0.4, which are almost unaffected by the next, very broad $d_{3/2}$ resonance at about 40 MeV.

Apart from the change in overall pairing intensity, see Table IV, norms of the lower HFB components $N_n$ shown in Fig. 10 closely follow the shape of localizations. In order to further visualize this property, in Figs. 11 and 12 we show results corresponding to the $2d_{3/2}$ resonances when they are moved up to the top of the centrifugal barrier. Six panels presented in the Figures have been obtained by using the depth parameters $\nu_{d_{\frac{3}{2}}}$=4.9(0.1)4.4, which corresponds to shifting the bottom of the $d_{3/2}$ central potential well from $-48.4$ to $-39.8$ MeV. As a result, the $2d_{3/2}$ resonances move from about 0.9 MeV to 6 MeV (at the same time the barrier also slightly increases, from 5 to 6 MeV). Hence, the top panels in Figs. 11 and 12 correspond to a broad resonance located right at the top of the centrifugal barrier.

Localizations of the PTG' $d_{3/2}$ continuum states (Fig. 11) closely follow the pattern of broadening and rising resonances. Apart from the lowest two panels, corresponding to low-energy resonances, the HFB localizations (dots) do not differ from the PTG' results (lines). This is because the larger the distance of the resonance from the Fermi surface ($\lambda_N \simeq -0.4$ keV in all cases), the weaker are, of course, modifications induced by pairing correlations. Moreover, the rising resonance leaves behind (at low energies) very low values of localizations, which additionally contributes to a decreasing of the overall pairing intensities.

Interestingly, norms of the lower HFB components $N_n$, presented in Fig. 12, closely follow the pattern of localizations. It means that values of localizations of the PTG' single-particle states indeed decide about the pairing coupling of the HFB quasiparticle continuum states, cf. Eq. (5.6). However, again none of the quasiparticle states (close to resonant energies or not) can be used as a single representative of the continuum phase space. Even for the resonances located very deeply inside the barrier (the lowest panels in Fig. 12), norms $N_n$ for quasiparticle states at resonance energies do not exhaust the occupation numbers $v_{\text{can}}$.

In Figs. 11 and 12, arrows indicate values of the canonical $2d_{3/2}$ energies $\epsilon_{\text{can}}$. With resonances moving up, these energies increase too, and appear always slightly above the resonant energy. These canonical states represent correctly the whole region of the quasiparticle continuum within, and above the centrifugal barrier. The $1d_{3/2}$ canonical states are deeply bound ($\sim -25$ MeV), and the $3d_{3/2}$ canonical states are very high ($\sim +25$ MeV), so indeed those corresponding to $2d_{3/2}$ (as shown in the Figures) are well separated representatives of a wide region of the phase space. Therefore, any approximate scheme, aiming at a correct description of pairing correlations near continuum, should vie to find these states at right places. Unfortunately, the HF+BCS method which uses resonant continuum states only, does not seem to have a chance to attain such a goal. However, similarity of localizations and norms $N_n$ found in the present study may give hope for approximate solutions of HFB equations. Of course, when the spherical symmetry is conserved, direct solution of the HFB problem is easy enough that no approximated methods are necessary. However, this is not the case for the deformed systems, where moreover, single-particle resonances split and become much more difficult to treat.

## VI. PERSPECTIVES AND OUTLOOK

The states at positive energies having maximum localization inside the region of potential well, preserve a remnant of the eigenvalue structure [68,69]. Whether virtual or resonant, these low lying states are likely to play an important role in the low-energy phenomena. For example, the neutron direct radiative capture process, which depends on properties of nuclear potential both at the surface and in the interior regions, is a unique tool to investigate properties of nuclear wave functions. Similar information may also be derived from the inverse process, i.e., the Coulomb dissociation process, and both those processes are going to be very important in extracting the information about nuclei close to the drip line. Raman *et al.* [70] have shown how strongly the details of a neutron-nucleus potential affect the capture mechanism for *s*-neutrons. (Extension of this analysis to higher partial waves has been done recently by Mengoni *et al.* [71].) This is also important in many astrophysical issues, e.g., the inhomogeneous Big Bang model or the neutron poison problem.

Low-lying resonant states may influence the pairing correlations enhancing the binding energy of halo nucleons (neutrons). In this work we have seen that a presence near threshold of the $L=0$ poles of the $S$-matrix corresponding to bound, virtual, and resonance states, may influence the pairing correlations and increase the localization of scattering wave functions in the narrow range of $\Delta\epsilon \simeq 2$–$4$ MeV above the threshold. Hence, the pairing interaction induces a subtle rearrangement effect in the structure of scattering states which modifies the strength of pairing field and the binding energy of affected nucleons close to the Fermi surface what, in turn induces new rearrangement in the structure of low lying scattering states, etc. This effect can be described only in the self-consistent treatment of pairing correlations.

For the HF potentials where the resonance width $\Gamma$ tends to a finite value when the resonance energy $\epsilon$ tends to zero, in the physical situation corresponding to the drip lines, i.e., when the nucleon (neutron) separation energy tends to zero ($S_n \to 0$), and hence $-\lambda_n - \Delta \simeq 0$, the standard quasiparticle picture [72] may be questionable. For the quasiparticle picture to be valid, the appreciable fraction of the single-particle strength should remain in the quasiparticle excitations and, moreover, the lifetime of excitations about the Fermi surface should be relatively long. Whether the single-particle strength is concentrated in the quasiparticle excitations depends on the strength of the residual interaction mixing the single-particle strength into more complicated configurations. But even when the residual interaction is strong, one can always assure in normal systems that the quasiparticle lifetime is long, i.e., the spreading of the quasiparticle strength function is small. In those systems, width of a quasiparticle state $\Gamma(k) \sim (k - k_F)^2$ can be made small by letting $k$ approach $k_F$ [73]. This result is independent of details of the interaction and follows from the Pauli principle and the phase-space considerations. However, as demonstrated in the example of the PTG potential, the states above the Fermi surface of $k_F \simeq 0$ may have a finite width in the limit of $k \to k_F$. In spite of the fact that the PTG centrifugal barrier is different from the physical centrifugal barrier, this feature should be seen in the structure of weakly bound nuclei near the drip line, at least for systems where $L=0$ states are present near threshold.

This particular aspect does not seem to be modified by including the pairing correlations within the HFB method. Consequently, the perturbation theory may not be applicable in these systems, and various kinds of instabilities caused by the residual correlations are expected. These instabilities may change the initial HF(B) vacuum in these nuclei non-perturbatively. In other words, one may expect that the spectrum of excitations in the unperturbed HF(B) system may not be compatible with the spectrum of excitations in the HF(B) system perturbed by the residual correlations. This brings about a new and challenging aspect into the experimental studies of shell structure and excitations close to the drip line: The atomic nuclei at the drip-line may provide a new kind of quantum open system with yet unknown properties of its excitation spectrum, resulting from the strong coupling between bound interior states and the environment of scattering states. If discovered, its microscopic description may call for new techniques in solving the quantum many-body problem. Similar coupling between the localized quasi-stationary states and the scattering states has already been recognized to be responsible for the unusual features of the surface of nuclear potential [58].

One of the most interesting aspects of correlations in drip-line systems is the question of multipole instabilities and deformations. A basis to study such phenomena should be provided by including the HFB pairing correlations in deformed states. This is a very difficult task, and only recently first attempts of such solutions become available [17,18,34]. The problem here is related to solving the HFB equations in a deformed coordinate-space representation, in a situation where variational methods are not applicable because the HFB equation has a spectrum unbounded from below. However, in our opinion, only such an approach may provide a sound basis of quasiparticle states in which other correlations (perturbative or not) may further be investigated.

It is essential to distinguish between five different aspects of the pairing coupling to the continuum phase space. First, enough continuum has to be included to cover the zone around the Fermi level, which is reasonably larger than the pairing gap. For gaps of about 1 MeV the zone which is 5 MeV wide is often used, cf. Ref. [17]. Second, the low-lying continuum with large localizations has to be taken into account. In cases studied here (Fig. 5), the zone of 5 MeV seems to be enough, however, this aspect strongly depends on the shape and depth of the single-particle potential, and the safe limit should probably be at least twice larger. Third, enough continuum should be included so as the contributions $N_n$ to particle number become small, and quasiparticle states entering and leaving this zone do not cause significant changes. A limit of $N_n \simeq 0.001$–$0.0001$ is probably the least one can get away with, which already requires going up to 10-15 MeV into the continuum, see Fig. 8. Fourth, all quasiparticle states which significantly contribute to the canonical states (i.e., those which have large spectral amplitudes [12]) should be included, which requires 10-20 MeV of the continuum [12]. Finally, a coupling between particle-type and hole-type quasiparticle states has to be taken into account. Although this coupling is not responsible for the physical widths of deep hole states, cf. discussion in Ref. [12], it is present in the HFB equations, and affects continuum solutions. This aspect requires taking continuum up to energies exceeding the depth of the single-particle potential, i.e., to about 40-50 MeV, and such a prescription [6] has been used here. Needless to say that the above discussion concerns the HFB continuum;

analyses based on the BCS method, which use the single-particle continuum, have other properties, and often diverge with increasing continuum cut-off energy.

## VII. CONCLUSIONS

Much is known about the analytic properties of one-particle wave functions in the low energy continuous spectrum when there is a real, virtual or quasistationary level with energy close to zero [68,69]. It was shown for a number of potentials, that the continuum wave functions in a wide range of **r**-values have the **r**-dependence which is remarkably close to that of the wave function for the zero-energy level [74,68,69]. Validity of this finding has been also discussed in Sec. IV B (see Fig. 3). As noticed by Migdal *et al.* [68], this approximate factorization of the continuum wave functions in the resonance region of one-particle phase space should simplify the computation of matrix elements. In the HF+BCS approach of Refs. [30,31], it is assumed that the analytical features of the $S$-matrix, which lead to the factorization property of the single-particle wave functions in the resonance region, remain valid in the presence of pairing correlations. Actually, this weak perturbation hypothesis for the $S$-matrix in the presence of pairing interaction remains not proved. To which extend the structure of single-particle resonances and the non-resonant continuum is affected by the presence of short-ranged correlations of the pairing type is the main problem which we have addressed in this work using the two-component quasiparticle wave-functions of the HFB approach and the PTG single-particle potential for which the $S$-matrix properties are known analytically. The detailed analysis performed in this work, shows that this weak perturbation assumption for the analytic structure of the $S$-matrix for the one-particle problem may be hazardous in many situations.

The comparison of the localization of HFB upper component of the quasiparticle states with the localization of the $s_{1/2}$ continuum PTG states exposed a complicate interplay between resonant and non-resonant HFB continuum which by no means can be approximated by the HF+BCS approximation spanned on the skeleton of the $S$-matrix resonances for the one-particle problem. The energy position of canonical states, which govern the pairing properties of the system, is not correlated with the presence and position of pole in the $S$-matrix for the one-particle problem. As a consequence, the pairing coupling of resonant and non-resonant $s_{1/2}$ continuum is quite similar, and its magnitude depends on the single-particle localization in the interval of 2–3 MeV above the Fermi surface.

When the single-particle poles are very close to the real axis in the momentum space ($\Re(k) > 0$), like in the case of very narrow high-$j$ resonances deep inside the centrifugal barrier, the pairing interaction is too weak to perturb the single-particle pole structure and, hence, these resonances are not very different from the bound states.

In the intermediate case, like $d_{3/2}$-resonances studied in Sec. V B, the situation depends on the position of the single-particle resonance with respect to the top of the centrifugal barrier. In a typical case, however, norms of the lower HFB components closely follow the pattern of localizations in the corresponding single-particle continuum, i.e., localizations of the single-particle continuum states determine the strength of the pairing coupling of the HFB quasiparticle continuum. However, like for the $s_{1/2}$ continuum, none of the quasiparticle states can be used a as a single representative of the continuum phase space.

## ACKNOWLEDGMENTS

This research was supported in part by the Polish Committee for Scientific Research (KBN) under Contract No. 2 P03B 040 14, by the U.S. Department of Energy under Contract Nos. DE-FG02-96ER40963 (University of Tennessee), DE-FG05-87ER40361 (Joint Institute for Heavy Ion Research), DE-AC05-96OR22464 with Lockheed Martin Energy Research Corp. (Oak Ridge National Laboratory), by the NATO grant CRG970196, by the French-Polish integrated actions programme POLONIUM, and by the computational grant from the Interdisciplinary Centre for Mathematical and Computational Modeling (ICM) of the Warsaw University. The authors wish to thank W. Nazarewicz for the hospitality extended to them during the visit at ORNL where part of this work has been done and for the critical reading of the manuscript.

TABLES

TABLE I. Parameters used to draw the potentials presented in Fig. 1. Parameters $\nu_{g\frac{9}{2}}$ and $\nu'_{g\frac{9}{2}}$ are used with the PTG and PTG' centrifugal barriers, respectively.

| $\Lambda$ | $s[\text{fm}^{-1}]$ | $\nu_{s\frac{1}{2}}$ | $\nu_{g\frac{9}{2}}$ | $\nu'_{g\frac{9}{2}}$ |
|---|---|---|---|---|
| 1 | 0.2192 | 6.719 | 6.188 | 7.199 |
| 3 | 0.09259 | 5.165 | 5.470 | 5.176 |
| 7 | 0.04059 | 5.034 | 5.372 | 4.693 |

TABLE II. Depth parameters of the PTG potential, for which the $3s_{1/2}$ state is resonant, virtual, or bound, and for which the $2d_{3/2}$ state is resonant or virtual. The other two parameters are $\Lambda=7$ and $s=0.04059$.

| case | | $\nu_{Lj}$ | energy |
|---|---|---|---|
| (a) | resonant $3s_{1/2}$ state | 4.882 | $\epsilon_{\text{res}}=(74-4225i)$ keV |
| (b) | virtual $3s_{1/2}$ state | 4.972 | $\epsilon_{\text{virt}}=-84$ keV |
| (c) | bound $3s_{1/2}$ state | 5.034 | $\epsilon_{\text{bound}}=-74$ keV |
| (d) | resonant $2d_{3/2}$ state | 4.850 | $\epsilon_{\text{res}}=(649-2610i)$ keV |
| (c) | virtual $2d_{3/2}$ state | 4.900 | $\epsilon_{\text{virt}}=(-247-1776i)$ keV |

TABLE III. Depth parameters of the PTG' potential (2.9) defining the neutron spectrum used in the HFB calculations. Parameters not given here are listed in Table II.

| $Lj$ | $\nu_{Lj}$ | $Lj$ | $\nu_{Lj}$ | $Lj$ | $\nu_{Lj}$ | $Lj$ | $\nu_{Lj}$ |
|---|---|---|---|---|---|---|---|
| $s_{1/2}$ | Table II | $p_{1/2}$ | 4.640 | $p_{3/2}$ | 4.880 | $d_{3/2}$ | Table II |
| $d_{5/2}$ | 5.180 | $f_{5/2}$ | 4.300 | $f_{7/2}$ | 4.720 | $g_{7/2}$ | 4.420 |
| $g_{9/2}$ | 4.800 | $h_{9/2}$ | 4.493 | $h_{11/2}$ | 4.992 | $i_{11/2}$ | 4.493 |
| $i_{13/2}$ | 4.493 | $j_{13/2}$ | 4.493 | | | | |

TABLE IV. Properties of the HFB solutions obtained for the PTG' potentials which give the resonant (a), virtual (b), and weakly bound (c) $3s_{1/2}$ single-particle states, or the $2d_{3/2}$ low-energy narrow resonances at two different positions. The neutron Fermi energies $\lambda_N$ and average pairing gaps $\langle\Delta_N\rangle$ [6] are given together with the canonical energies $\epsilon_{\text{can}}$ [12] and occupation factors $v^2_{\text{can}}$. All energies are in keV.

| case | $\lambda_N$ | $\langle\Delta_N\rangle$ | $Lj$ | $\epsilon_{\text{can}}$ | $v^2_{\text{can}}$ |
|---|---|---|---|---|---|
| (a) | $-314$ | 1146 | $3s_{1/2}$ | 2148 | 0.0321 |
| (b) | $-384$ | 1292 | | 1043 | 0.1002 |
| (c) | $-488$ | 1421 | | 390 | 0.2083 |
| (d) | $-436$ | 1217 | $2d_{3/2}$ | 1765 | 0.0548 |
| (c) | $-488$ | 1421 | | 1107 | 0.1149 |

FIGURES

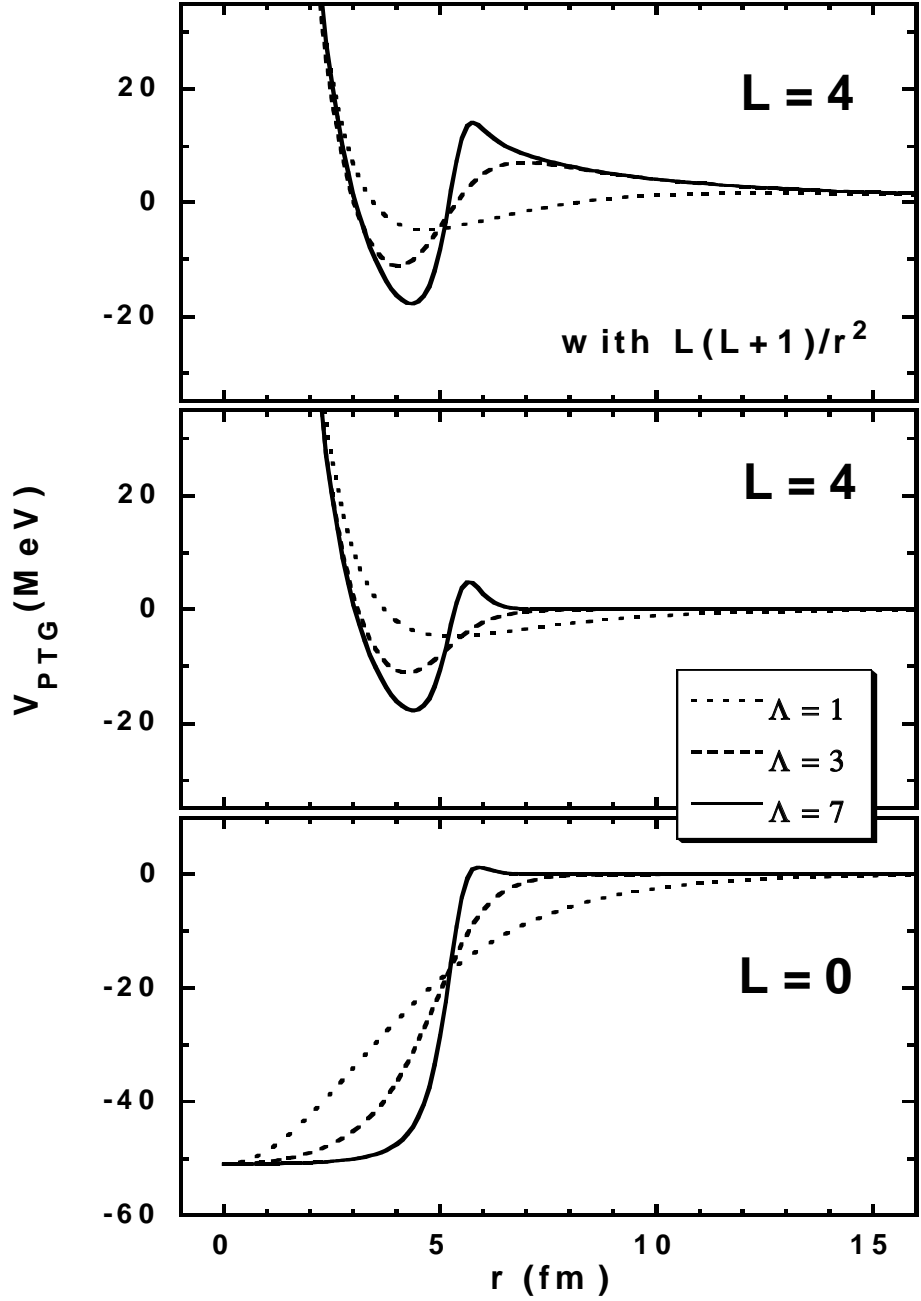

FIG. 1. Radial PTG potentials for three values of the parameter $\Lambda$ (see Table VII for the complete set of parameters). The bottom panel shows the potential for $L=0$, while the middle and top panels correspond to $L=4$ and the PTG (2.1) and PTG' (2.9) potentials, respectively (see text).

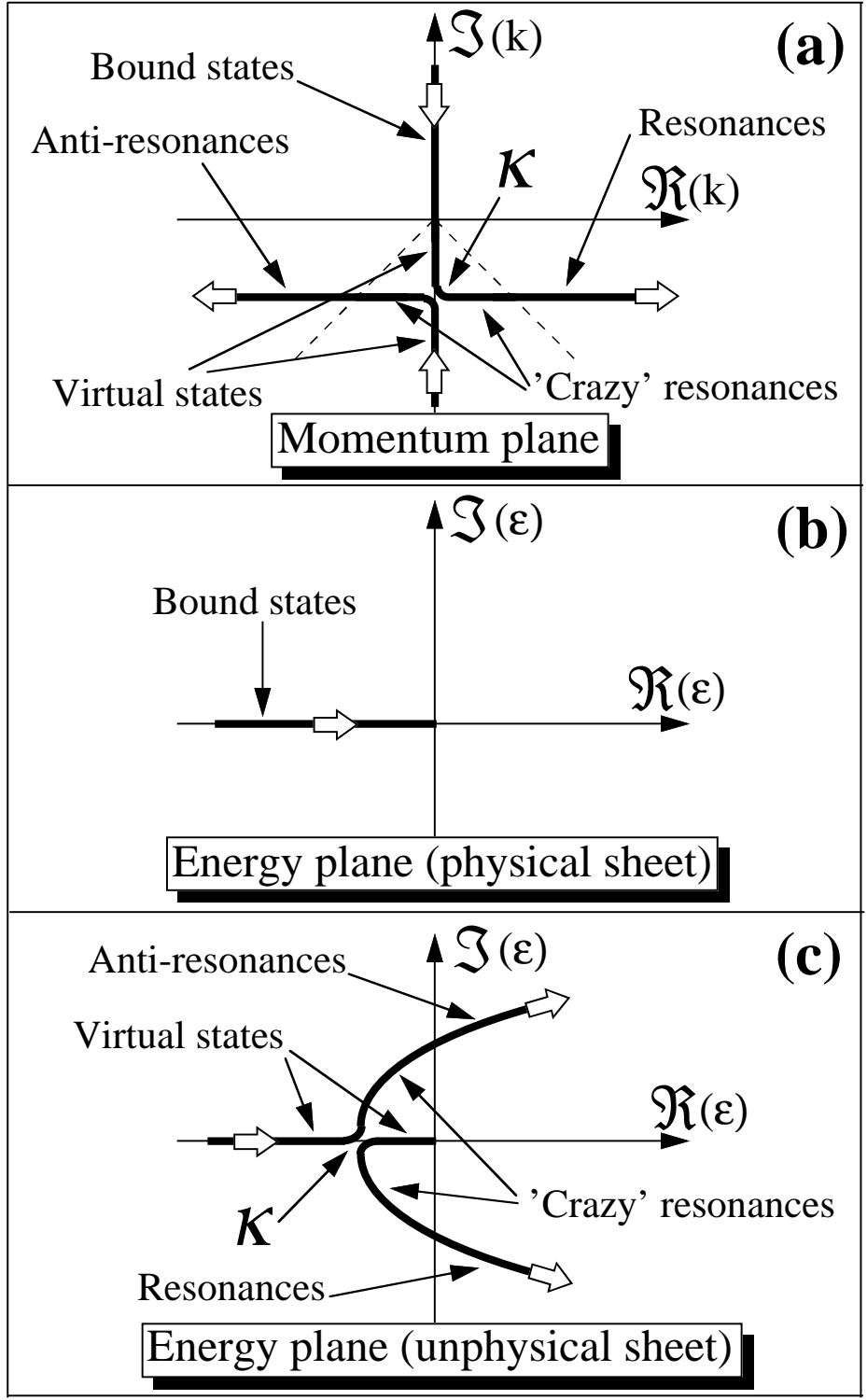

FIG. 2. Schematic representation of different domains of $S$-matrix poles in the complex momentum and energy planes.

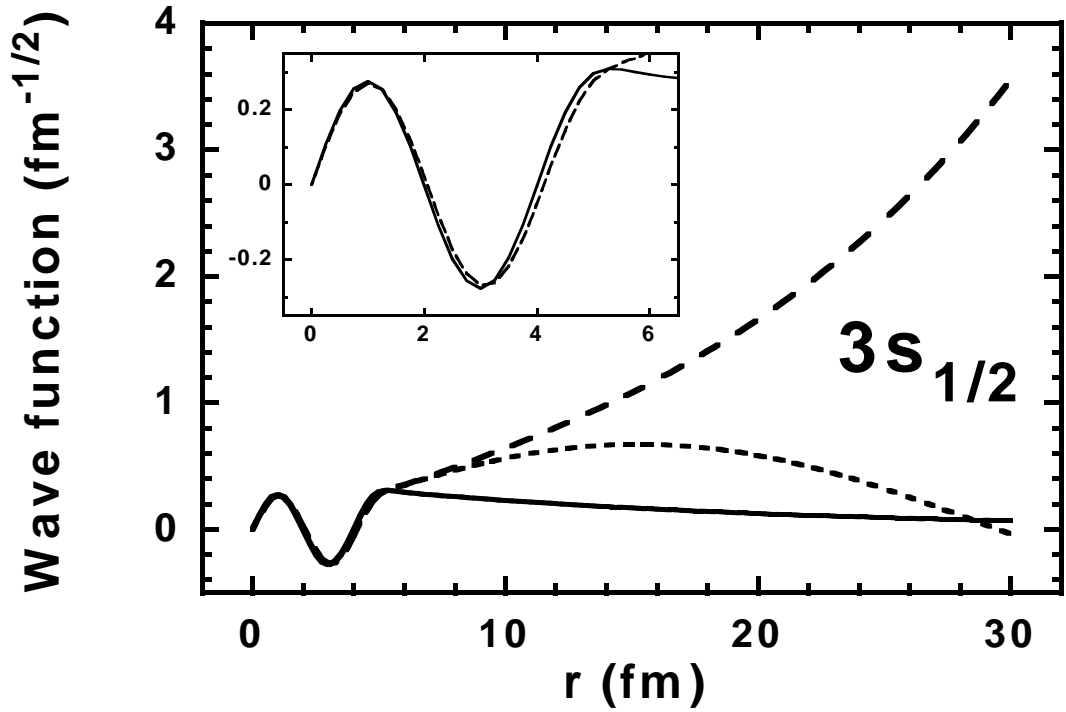

FIG. 3. Radial PTG wave functions of the $3s_{1/2}$ state calculated analytically for parameters listed in Table II. The dotted, dashed, and solid lines correspond to resonant, virtual, and bound $3s_{1/2}$ states, respectively. In the inset the same wave functions are shown in the region of attractive potential. Volume element $4\pi r^2$ is included in the definition of wave functions.

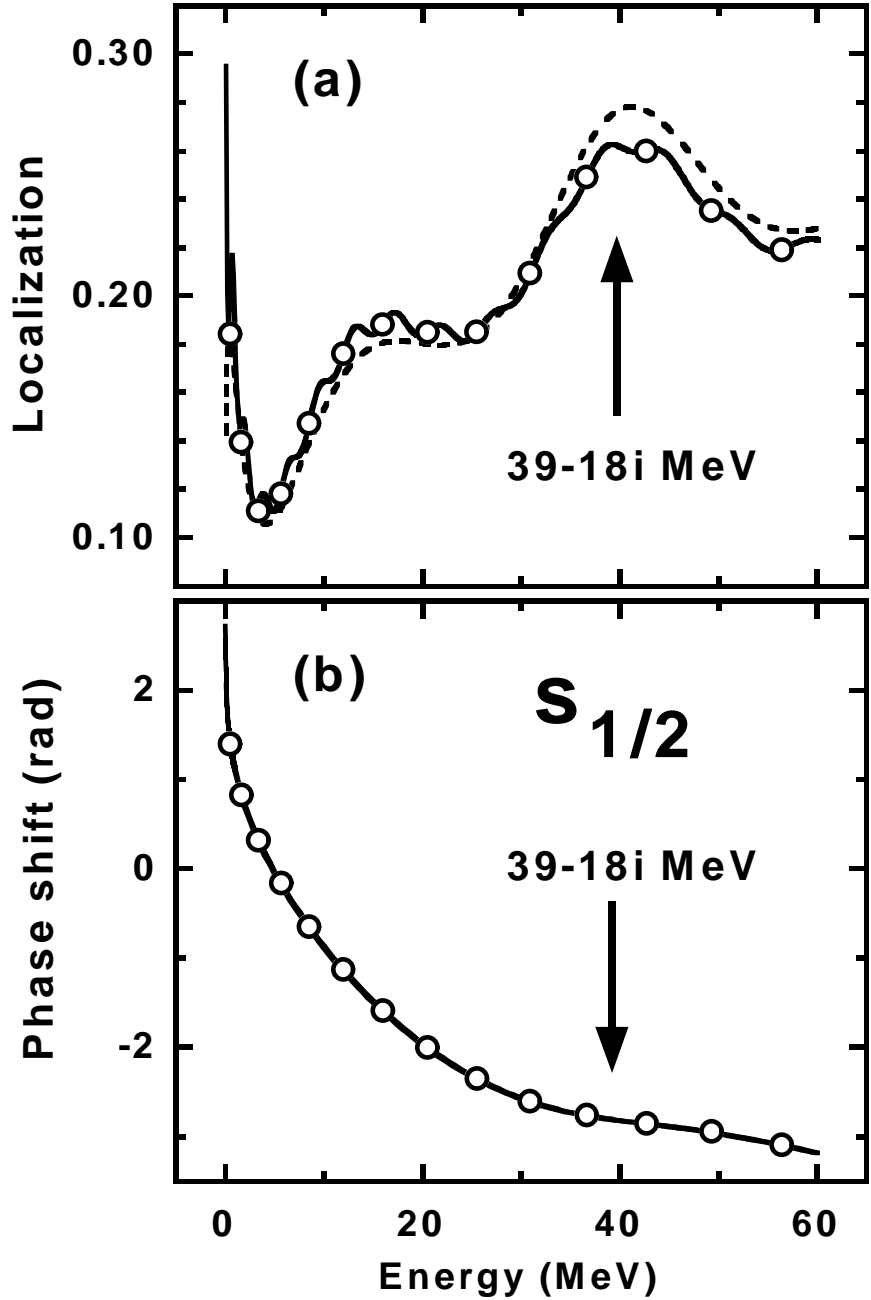

FIG. 4. Localization [Eq. (4.19)] (a) and phase shift (b) of the $s_{1/2}$ continuum PTG states. Circles denote results of numerical calculations performed in the box of 30 fm. Solid and dashed lines show the localizations calculated analytically for scattering wave functions normalized in the box of 30 fm, and to a constant amplitude of the outgoing wave, respectively. Arrows indicate position of the lowest, very broad, resonance. Parameters of the PTG potential are given in Table II (bound $3s_{1/2}$ state).

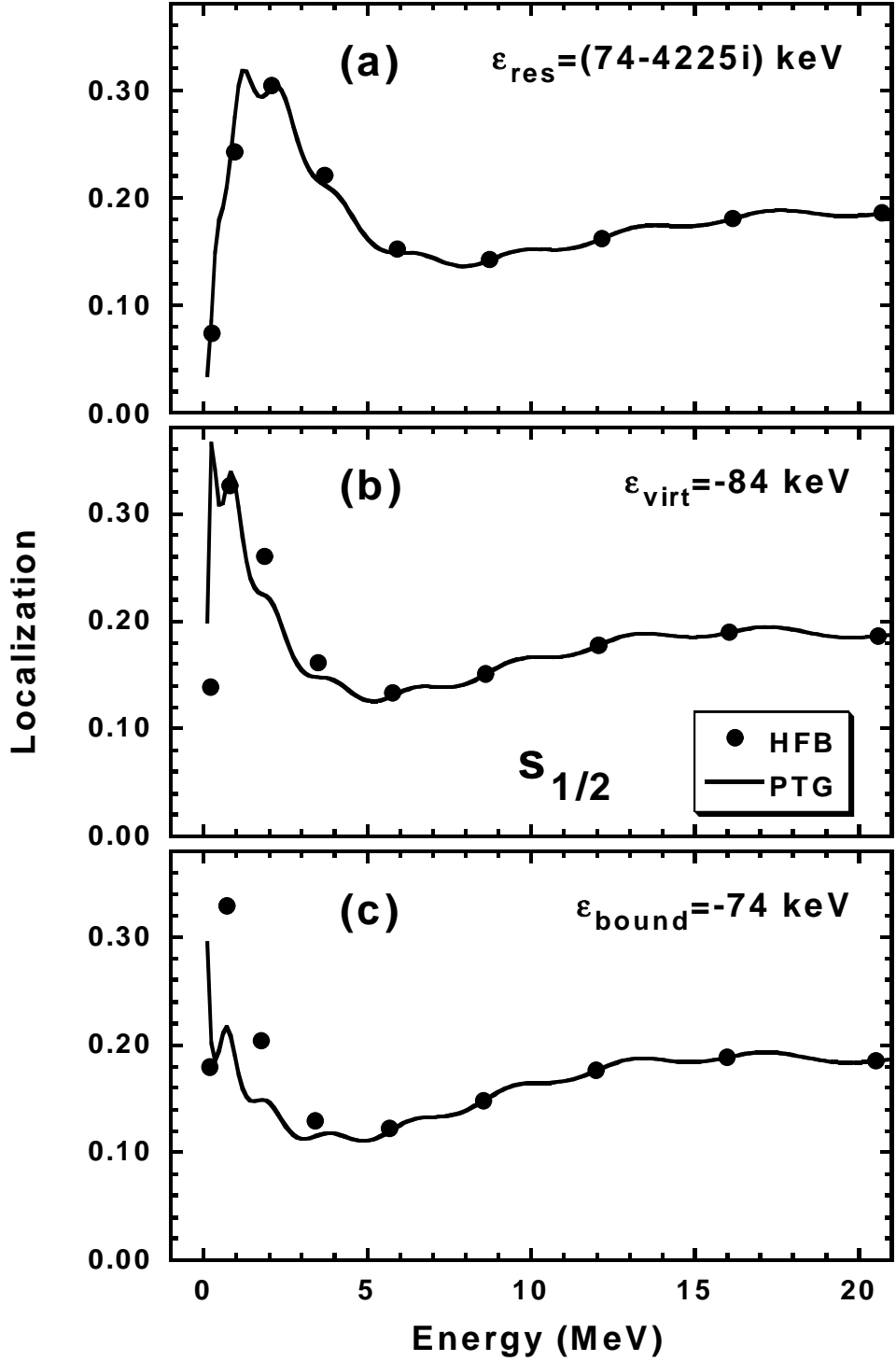

FIG. 5. Localization [Eq. (4.19)] of the HFB upper components (dots) of the quasiparticle states, compared with the localization of the $s_{1/2}$ continuum PTG states (lines).

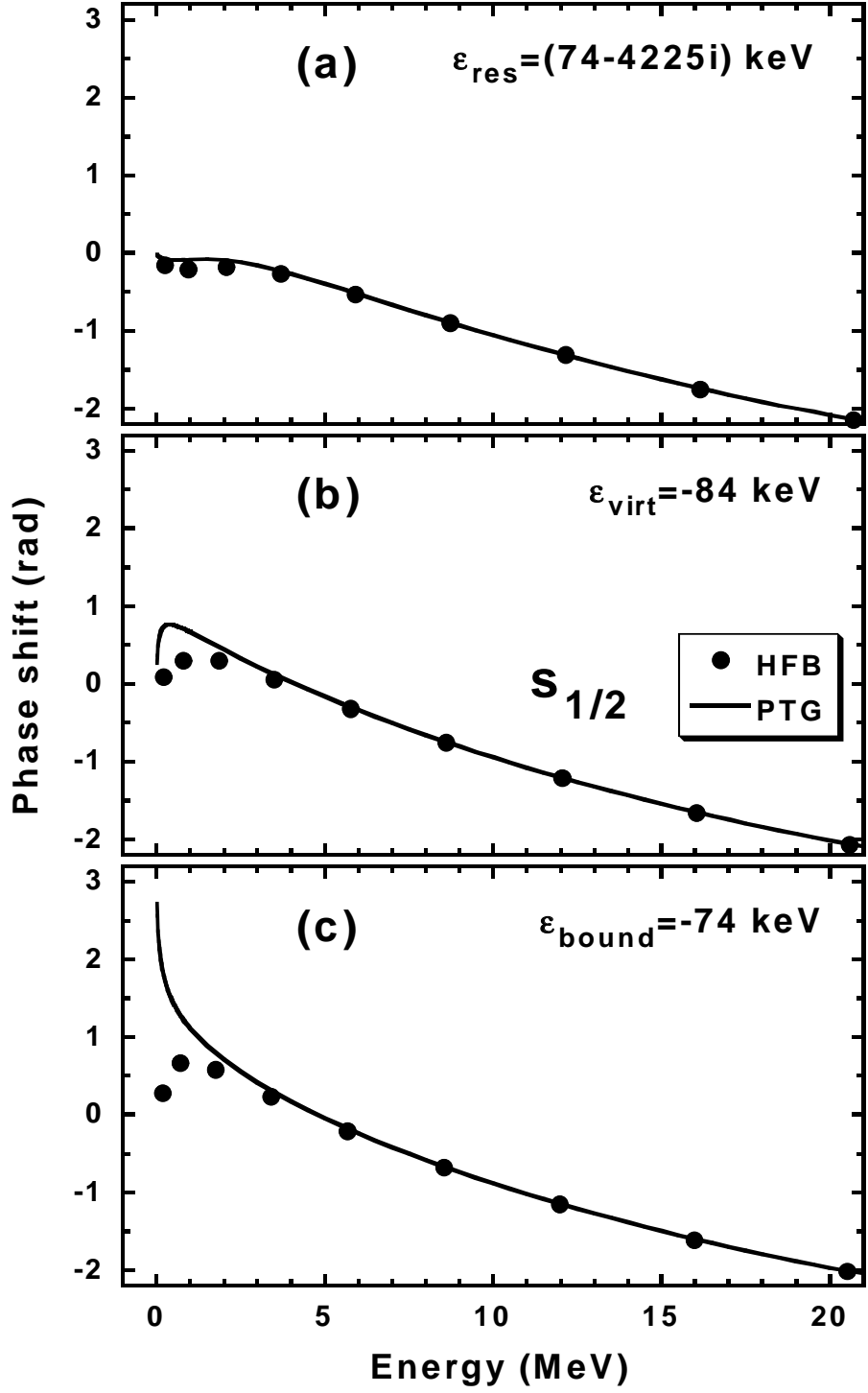

FIG. 6. Phase shifts of the HFB upper components (dots) of the quasiparticle states, compared with the phase shifts of the $s_{1/2}$ continuum PTG states (lines).

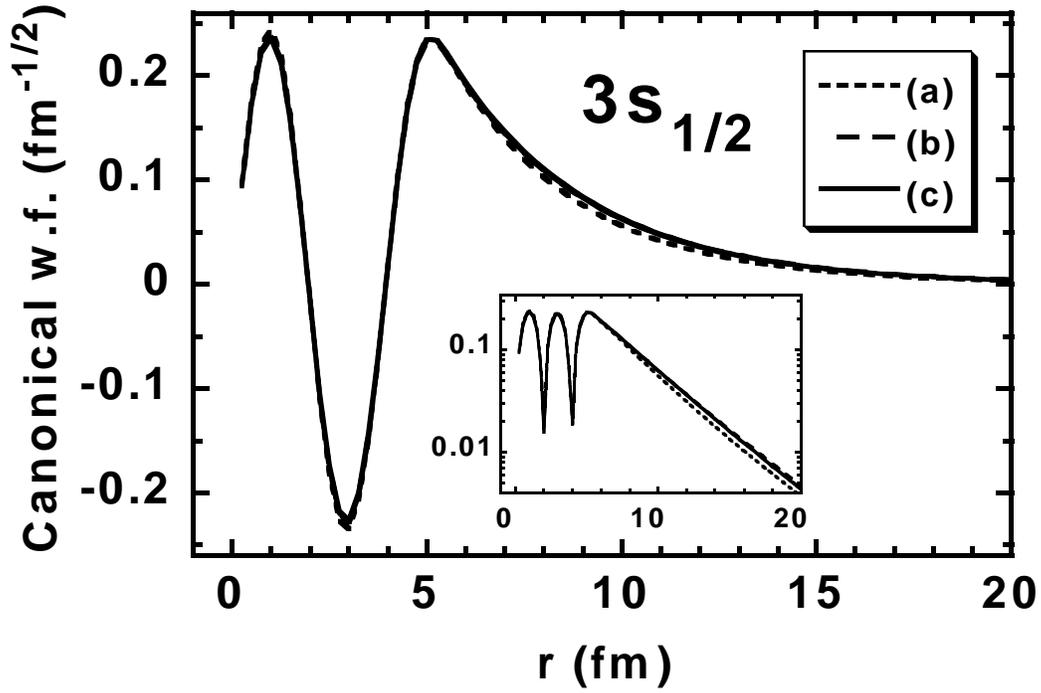

FIG. 7. Canonical wave functions of the $3s_{1/2}$ state, corresponding to the three cases of the HFB calculations listed in Table IV. The dotted, dashed, and solid lines correspond to resonant, virtual, and bound PTG $3s_{1/2}$ states, respectively. In the inset the same wave functions are shown in the logarithmic scale. Volume element $4\pi r^2$ is included in the definition of wave functions.

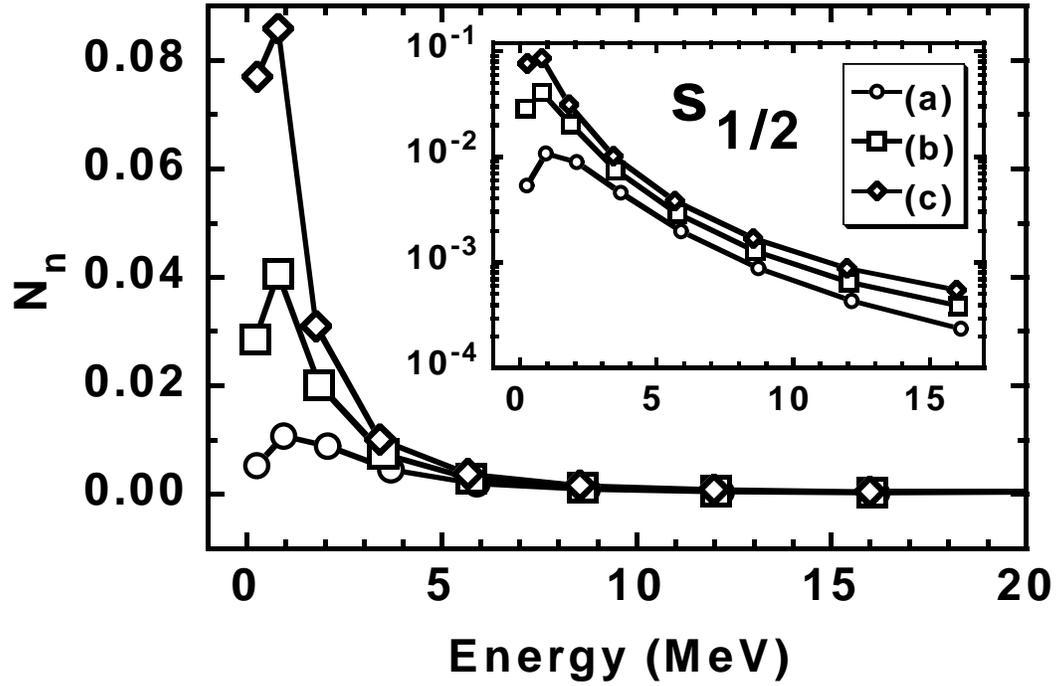

FIG. 8. Norms $N_n$ of the $s_{1/2}$ lower quasiparticle wave functions $\phi_2(E_n, \boldsymbol{r}\sigma)$, corresponding to the three cases of the HFB calculations listed in Table IV. Inset shows the same values in the logarithmic scale.

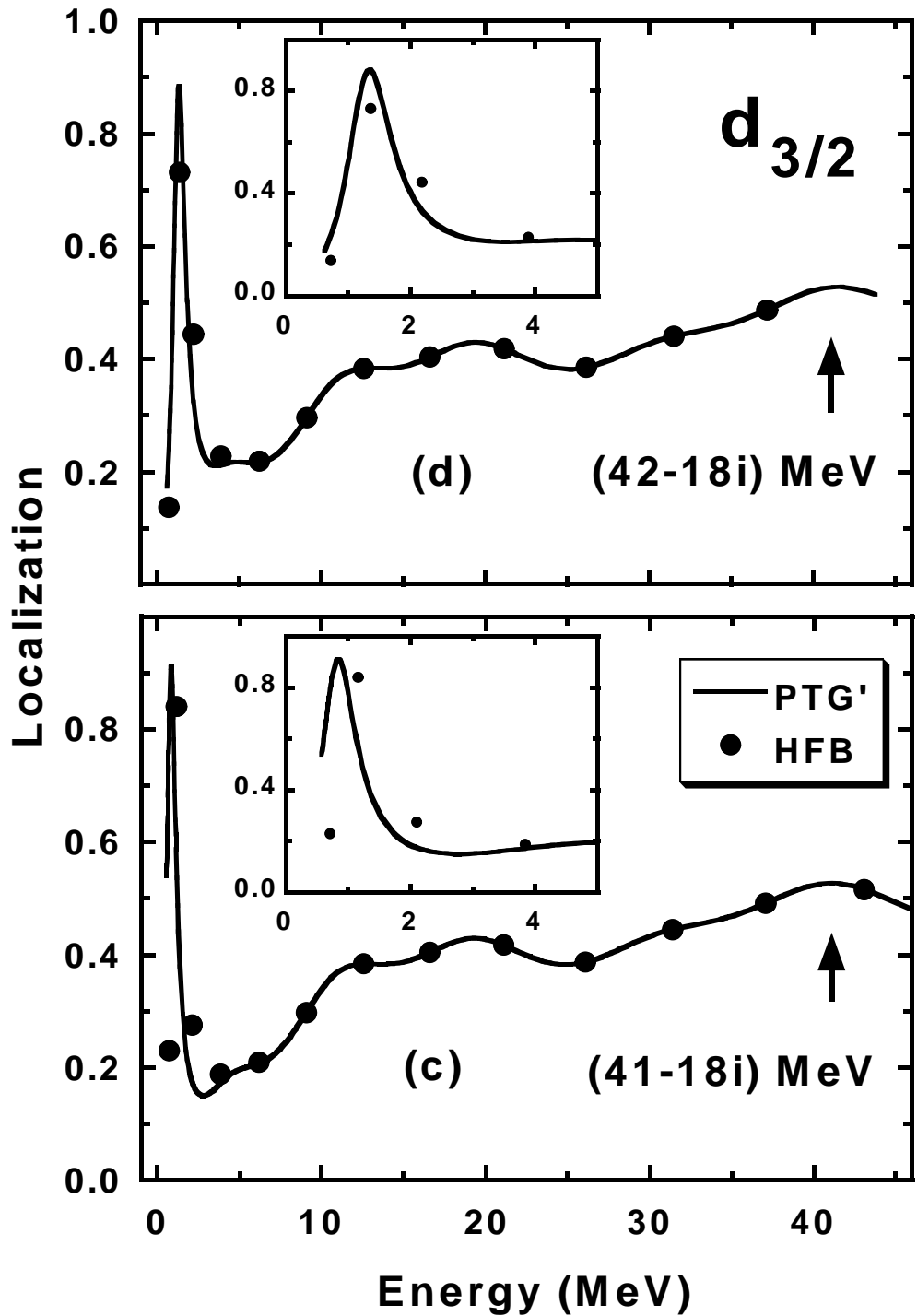

FIG. 9. Localization of the HFB upper components (dots) of the quasiparticle states compared with the localization of the $d_{3/2}$ continuum PTG' states (lines). Upper (a) and lower (b) panels differ by the position of the $2d_{3/2}$ resonances with respect to the top of the centrifugal barrier, see text.

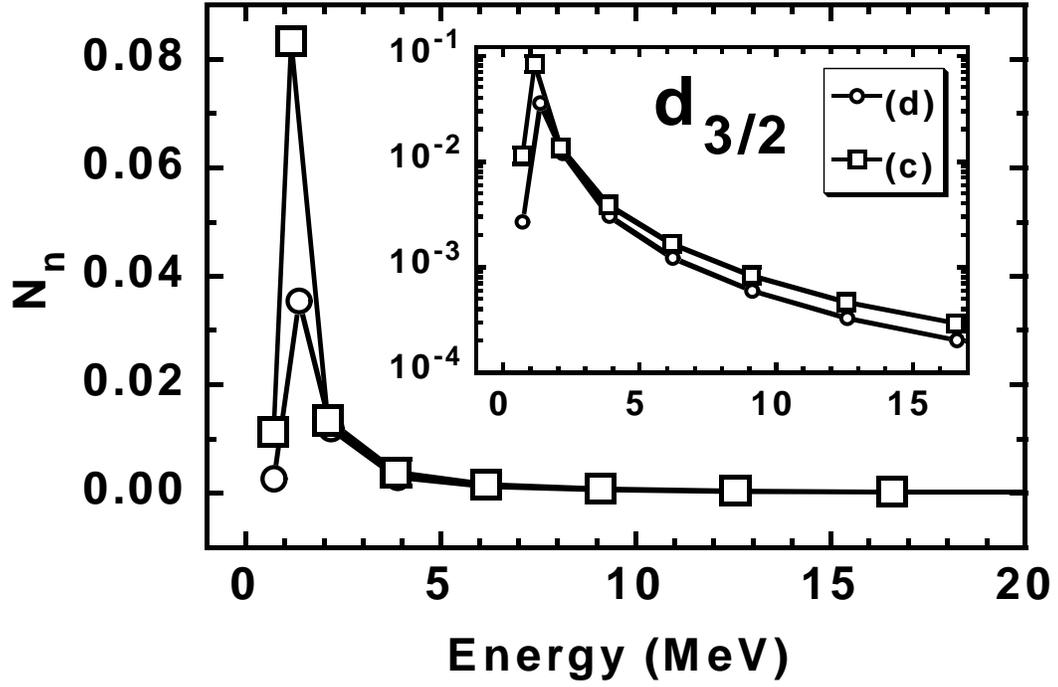

FIG. 10. Norms $N_n$ of the $d_{3/2}$ lower quasiparticle wave functions $\phi_2(E_n, \boldsymbol{r}\sigma)$, corresponding to the two cases of the HFB calculations listed in Table IV. Inset shows the same values in the logarithmic scale.

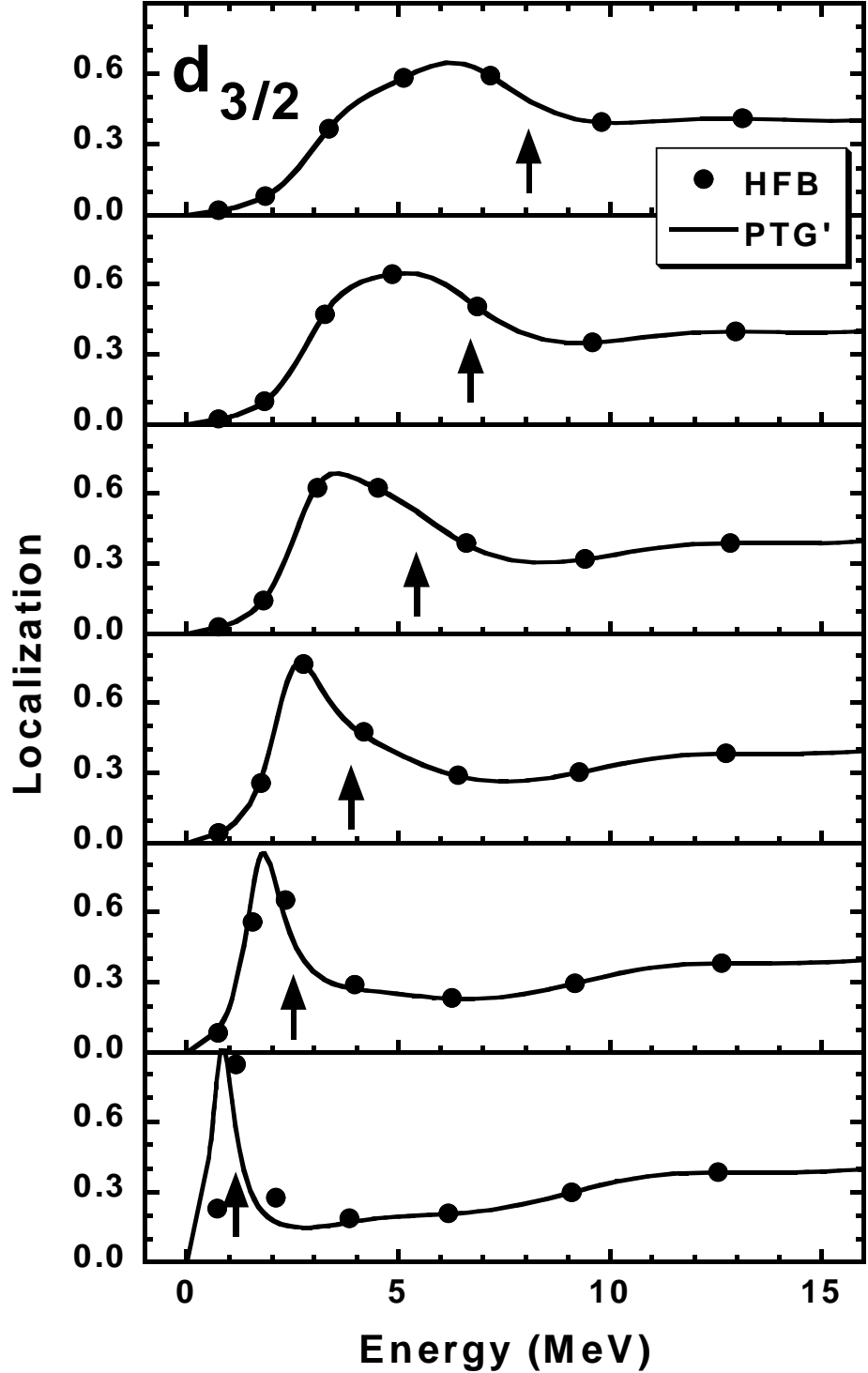

FIG. 11. Localizations of the HFB upper components of the $d_{3/2}$ quasiparticle states (dots) compared with those of the $d_{3/2}$ continuum PTG' states (lines). The six panels differ by the positions of the lowest $d_{3/2}$ resonance with respect to the top of the centrifugal barrier, see text. Arrows indicate values of the canonical energies $\epsilon_{\rm can}$ of the $2d_{3/2}$ canonical states.

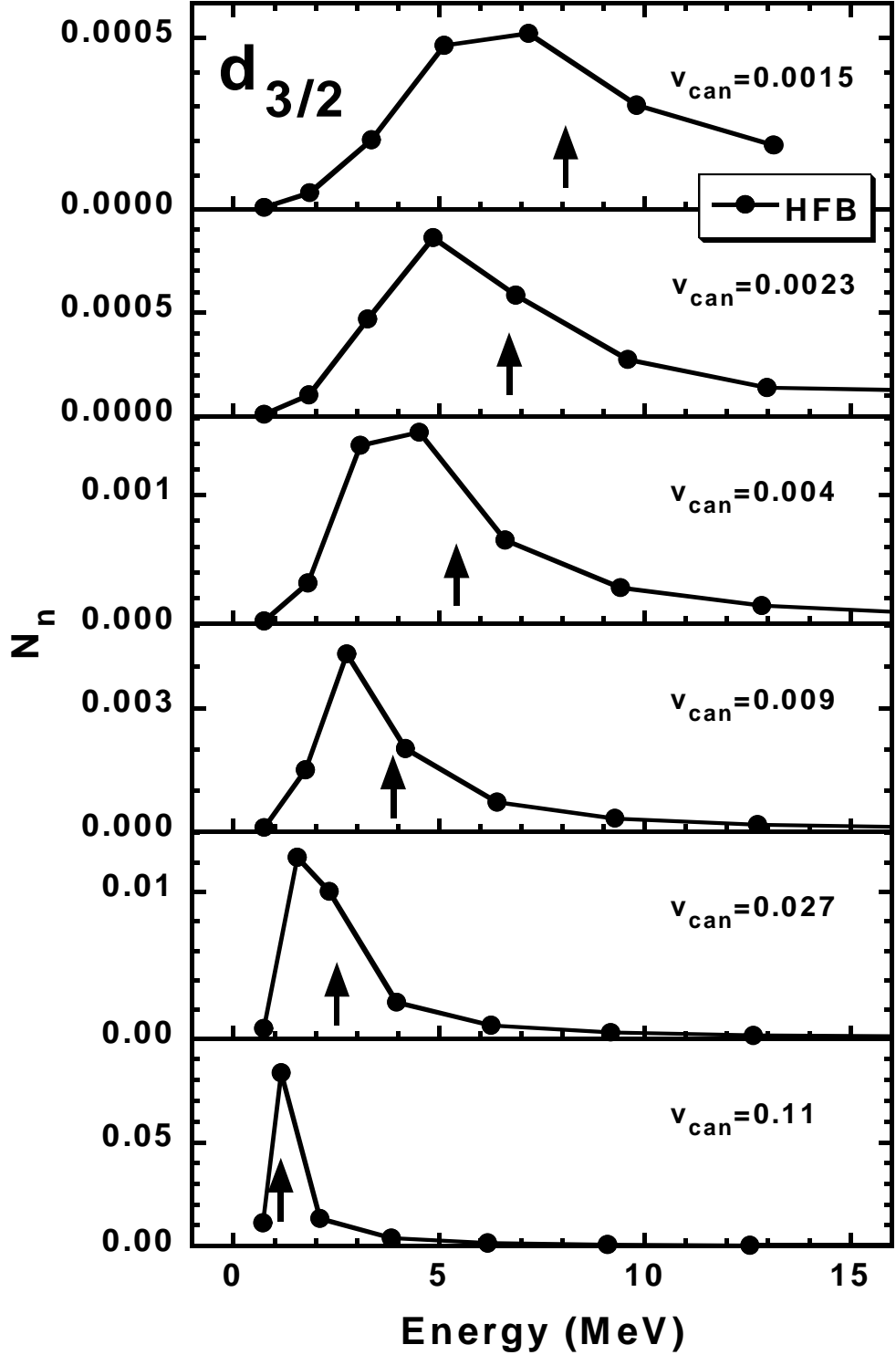

FIG. 12. Norms $N_n$ of the HFB lower components $\phi_2(E_n, \bm{r}\sigma)$ of the $d_{3/2}$ quasiparticle states. In each panel (similarly as in Fig. 11), arrows indicate values of the canonical energies $\epsilon_{\text{can}}$ of the $2d_{3/2}$ canonical states, while the corresponding occupation factors $v_{\text{can}}$ are given explicitly.